\documentclass[journal]{IEEEtran}

\usepackage{booktabs}
\usepackage{threeparttable}
\usepackage{amsmath,graphicx,cite,amssymb}
\usepackage{amsfonts,multirow,bm,array,setspace}

\usepackage{graphicx,float,cite,amssymb}
\usepackage{multirow,bm,array,setspace}
\usepackage{textcomp}
\usepackage{psfrag}
\usepackage{color}
\usepackage{pstricks,enumerate}
\usepackage{bbm}

\usepackage{algorithm,algpseudocode}
\makeatletter
\newcommand{\distas}[1]{\mathbin{\overset{#1}{\kern\z@\sim}}}%
\newsavebox{\mybox}\newsavebox{\mysim}
\newcommand{\distras}[1]{%
  \savebox{\mybox}{\hbox{\kern1pt$\scriptstyle#1$\kern1pt}}%
  \savebox{\mysim}{\hbox{$\sim$}}%
  \mathbin{\overset{#1}{\kern\z@\resizebox{\wd\mybox}{\ht\mysim}{$\sim$}}}%
}
\makeatother
\ifCLASSINFOpdf
\else
\fi
\newtheorem{proposition}{Proposition}

\newtheorem{remark}{Remark}

\newcommand{\hf}{\hat f}

\newcommand{\hF}{\hat F}

\newcommand{\hQ}{\hat Q}
\newcommand{\hhQ}{\hat{\hat Q}}

\begin{document}
%\title{Full-Duplex Cellular System with Cross-Channel Transmission and Interference Cancellation}
%\title{Power Adaptive HARQ for Ultrareliability via\\a Novel Outage Probability Bound and\\ Geometric Programming}
\title{Energy Efficient HARQ for Ultrareliability via\\ a Novel Outage Probability Bound and\\ Geometric Programming}
\author{%\IEEEauthorblockN{Kaiming Shen, \IEEEmembership{Member,~IEEE}, Wei Yu, \IEEEmembership{Fellow,~IEEE},\\ Xihan Chen, \IEEEmembership{Member,~IEEE}, and Saeed R. Khosravirad, \IEEEmembership{Member,~IEEE}} % <-this % stops a space
%\thanks{Manuscript received \today. This paper has been presented in part at IEEE ICC 2021. 
%The work of K. Shen was supported in part by National Natural Science Foundation of China (NSFC) under Grant 62001411 and in part by a gift funding from Nokia Bell-Labs, the work of W. Yu was supported by a gift funding from Nokia Bell-Labs.
\IEEEauthorblockN{Kaiming Shen, \IEEEmembership{Member,~IEEE},
Wei Yu, \IEEEmembership{Fellow,~IEEE},\\ Xihan Chen, \IEEEmembership{Member,~IEEE}, and Saeed R. Khosravirad, \IEEEmembership{Member,~IEEE}} % <-this % stops a space
\thanks{Manuscript accepted in IEEE Transactions on Wireless Communications. The work of Kaiming Shen was supported by the National Natural Science Foundation of China (NSFC) under
Grant 62001411.
The work of Wei Yu was supported by a Gift from Nokia. This paper has been presented in part at IEEE ICC 2021 \cite{shen21_ICC}. \emph{(Corresponding author: Kaiming Shen.)}

Kaiming Shen is with the School of Science and Engineering, The Chinese University of Hong Kong (Shenzhen), Shenzhen 518172, China (e-mail: shenkaiming@cuhk.edu.cn).

Wei Yu is with The Edward S. Rogers Sr. Department of Electrical and Computer Engineering, University of Toronto, Toronto, ON M5S 3G4, Canada (e-mail: weiyu@ece.utoronto.ca). 

Xihan Chen is with the College of Information Science and Electronic Engineering, Zhejiang University, Hangzhou 310027, China (e-mail: chenxihan@zju.edu.cn).

Saeed R. Khosravirad is with Nokia-Bell Labs, Murray Hill, NJ 07974 U.S.A. (e-mail: saeed.khosravirad@nokia-bell-labs.com).
}}

%Telephone: (800) 555--1212, Fax: (888) 555--1212}


\maketitle

\begin{abstract}
    Hybrid automatic repeat request (HARQ) is a key enabler for ultrareliable communications. This paper optimizes transmit power for the initial transmission and the subsequent retransmissions of HARQ with either incremental redundancy or Chase combining, aiming to minimize the expected energy consumption given the target outage probability and the target latency. The main challenge is due to the fact that the outage probability is a complicated function of the power variables which are nested in successive convolutions. The existing works mostly use a classic upper bound to approximate the outage probability by assuming unbounded transmit power, then convert the original problem to a geometric programming (GP) problem. In contrast, we propose a novel and much tighter upper bound by taking the practical power limit into consideration. The new bound and the resulting new GP method are further extended to a broader group of channel models with various fading, multiple antennas, and multiple receivers. As shown in simulations, the GP method based on the new bound significantly outperforms the existing strategies that either fix transmit power or optimize power by the classic bounding technique.
\end{abstract}

\section{Introduction}

\IEEEPARstart{U}{ltrareliable} communications refer to transmitting data with a target outage probability lower than $10^{-5}$, as compared to the traditional cellular systems typically with an outage probability around $10^{-2}$\cite{bennisPROC18}. This stringent performance standard is driven by a multitude of evolving applications of the Internet of Things (IoT), e.g., ultra high-definition (UHD) video \cite{3GPP}. This paper seeks an energy-efficient implementation of the hybrid automatic repeat request (HARQ) protocol to accommodate the ultrareliability requirements imposed in these applications.

The main question of concern is how to deliver a fixed-size message toward the intended receiver(s) at the minimum energy cost, while satisfying the target outage probability and the target latency. In particular, by incremental redundancy coding, the receiver can accumulate the mutual information in time, so the ultimate outage probability is affected by the transmit power of every round of HARQ transmission. The accumulation of mutual information poses a challenge to power optimization, because it results in the power variables being nested in successive convolutions inside the outage probability function which is difficult to analyze. The main idea of this paper is to isolate the power variables from the successive convolutions by using a novel upper bound to approximate the outage probability function. We end up with a geometric programming (GP) problem that can be solved efficiently. The proposed GP method can be justified in two respects. First, its solution is guaranteed to satisfy the outage probability and latency constraints of the original problem. Second, its performance approaches the global optimum if the signal-to-noise ratio (SNR) is sufficiently high. We remark that the proposed power control method is fundamentally different from the existing GP methods in \cite{Lee_TWC10,Larsson_TWC11,Su_TCOM11,Ge_IET15} that all rest upon the classic outage probability bound in \cite{Laneman_Globecom03,Laneman_IT03} by assuming unbounded transmit power. The new bound proposed in this paper is much tighter by taking the practical power constraint into account. Moreover, we discuss its extensions to a general parametric fading model, to scenarios with multiple antennas and with multiple receivers.

%As a distinguishing characteristic of this new bound, the proposed bound takes practical constraints on the power into account and hence is much tighter than the classic bound in \cite{Laneman_Globecom03,Laneman_IT03} that simply assumes unbounded transmit powers. The extensions for different fading models, for multiple antennas, and for multiple receivers are further discussed in the paper.

%approximate the outage probability properly, thereby 

%Our approach relies on a new upper bound of the outage probability to approximate the original problem in a geometric programming (GP) form. By contrast, the existing GP methods in \cite{Lee_TWC10,Larsson_TWC11,Su_TCOM11,Ge_IET15} are all based on a classic upper bound \cite{Laneman_Globecom03,Laneman_IT03} with high-power assumptions and thereby are prone to overestimation of the outage probability when the transmit powers are constrained. Furthermore, we extend the proposed outage probability bound to account for multiple-input single-output (MISO) channels. This extension is critical to the proposed cooperative HARQ scheme in the presence of multiple receivers.%, whereby the already successful receivers are used as relays to help the rest.

In the existing literature, ultrareliable communications have been examined from a variety of perspectives. Short packet coding is brought to the fore because of the significant impact it has on bit error rate and latency. Many works \cite{Sybis_VTC16,Wu_WCOML18,Shirva_MCOM18} in this area are empirically based, aimed at determining what types of codes (e.g., LDPC and polar codes) are most suited for ultrareliable low-latency communications (URLLC). Another group of works analyze the tail behavior of the extreme events in ultrareliable communications, typically by means of the extreme value theory \cite{Liu_CL18,Zhou_JSAC21} and the stochastic network calculus \cite{bennisPROC18,xie_PIMRC20}. For the system level design of ultrareliable communications, resource allocation has attracted considerable research interests, often in conjunction with newly emerging techniques of the next-generation networks, such as network slicing \cite{Quek_JSAC19}, non-orthogonal multiple access (NOMA) \cite{Dogan_JSTSP19,Wang_VTC17}, and massive multiple-input multiple-output (MIMO) \cite{Popovski_TCOM19}. The other aspects of ultrareliable communications that have been studied in the existing works include waveform design \cite{Eldessoki_WSA17} and random access \cite{Jacobsen_Globecom17}.

The paper is most closely related to a line of studies in \cite{Lee_TWC10,Larsson_TWC11,Su_TCOM11,Ge_IET15} that leverage the classic outage probability bound in \cite{Laneman_Globecom03,Laneman_IT03} to approximate the complicated outage probability function thereby relaxing the intractable power control problem of HARQ as a GP problem. The new bound proposed in this paper approximates the outage probability function more exactly. As a result, in comparison to the existing methods in \cite{Lee_TWC10,Larsson_TWC11,Su_TCOM11,Ge_IET15} based on the classic bound \cite{Laneman_Globecom03,Laneman_IT03}, the newly proposed GP method is less prone to overestimating outage probability---especially with low transmit power as widely seen in the IoT scenarios. The paper mainly consists of two parts: the first part focuses on constructing a tighter outage probability bound for the conventional link-level Rayleigh fading channel as considered in \cite{Lee_TWC10,Larsson_TWC11,Su_TCOM11,Ge_IET15}, while the second part concerns a generalization to the parametric fading (which encompasses Rayleigh fading and Rician fading as special cases), multi-antenna transmissions, and broadcast transmissions.

%While the first part of the paper focuses on the Rayleigh fading model as in the existing works \cite{Lee_TWC10,Larsson_TWC11,Su_TCOM11,Ge_IET15}, the second part considers a general Nakagami fading model (which encompasses Rayleigh fading and Rician fading as two special cases). Furthermore, we incorporate multiple antennas and multiple receivers into the proposed bound and GP method.

%Aside from ultrareliable communications, power control for HARQ is also motivated by \emph{green communications} in light of the carbon footprint concern.
In the existing literature on HARQ \cite{Zorzi_Globecom14,Berder_15,Vangelista_15,Eriksson_14,Ge_CL15,Zheng_JSAC14,Szczecinski_TCOM16}, 
although the problem formulations vary widely depending on which type of HARQ is used (i.e., type-I, Chase combining, and incremental redundancy), it is always of central importance to cast the outage probability in a form amenable to analysis and optimization. 
For type-I HARQ, i.e, when the previous packets are all discarded, 
\cite{Zheng_JSAC14} suggests simplifying the outage probability by means of a log-domain threshold approximation; this approximation is further developed in \cite{Ge_CL15} to account for type-II HARQ with Chase combining. Another way of approximating the outage probability for Chase combining is proposed in \cite{Berder_15} based on successive optimization.
For type-II HARQ, \cite{Eriksson_14,Szczecinski_TCOM16} prove that the optimal transmit power is increasing in each retransmission under certain assumptions. Furthermore, \cite{Zorzi_WCL14,Zorzi_TWC14,Kim_Globecom13} consider HARQ in the finite blocklength regime. These works all assume uniform transmit power across the different rounds of HARQ retransmission in order to apply the outage analysis for block-fading channels in \cite{Polyanskiy_TIT14}. For HARQ with finite blocklength coding, \cite{Zorzi_Globecom14,Vangelista_15} allow the transmit power to vary from block to block, but they both restrict the power control problem to only two blocks because of the optimization difficulty.

The remainder of the paper is organized as follows. Section \ref{sec:sys} describes the system model. Section \ref{sec:upper_bound} introduces the main result of this paper: a novel upper bound on outage probability. Based on this new bound, a GP power control method is devised for HARQ in Section \ref{sec:link_GP}. Section \ref{sec:multidim} further develops the above results to account for more general channel models. Section \ref{sec:simulations} shows the simulation results. Finally, Section \ref{sec:conclusion} concludes the paper.

Throughout the paper, we use $\mathrm{Pr}[\mathcal E]$ to denote the probability of the event $\mathcal E$, 
use $\mathcal{CN}(0,\sigma^2)$ to denote a zero mean complex Gaussian distribution with the variance $\sigma^2$, and use $*$ to denote the convolution operation. We use $\mathbb R_+$ to denote the set of all positive real numbers, and use $\mathbb C$ to denote the set of all complex numbers. For ease of reference, Table \ref{tab:notation} summarizes the definitions of the main variables in the paper.

%. Section \ref{sec:link:method} first proposes a new upper bound, then introduces a GP algorithm for power control along with performance analysis. Extension of the above results to a multi-link scenario is presented in Section \ref{sec:sys} along with a topological power control framework. Finally, Section \ref{sec:conclusion} concludes this work. The following notation is used throughout the paper. We use $\mathbb P[\mathcal E]$ to denote the probability of some even $\mathcal E$, $\mathcal{CN}(\mathbf 0,\mathbf K)$ the multivariate complex Gaussian distribution with zero mean and covariance matrix $\mathbf K$, $\chi^2(d)$ the chi-square distribution with $d$ degrees of freedom. Further, let $\mathbf C^{a\times b}$ be the $a\times b$ dimensional complex space, let $\bI_L$ be the $L\times L$ identity matrix, and let $[a:b]$ be the set of integers $\{a,a+1,\ldots,b-1,b\}$. A summary of the main variable symbols is provided in Table \ref{tab:notation}.

\begin{table}[t]
\small
\centering
\caption{\small List of Main Variables}
{\renewcommand{\arraystretch}{1.2}
\begin{tabular}{|ll|}
%\hline
\hline
Symbol &  Definition \\
\hline
$N$ & maximum number of blocks\\
$M$ & number of transmit antennas\\
%$n$ or $m$ & Block index\\
$E$ & expected energy consumption\\
$D$ & average number of transmissions\\
$S$ & pathloss-to-noise ratio\\
$P$ & maximum transmit power\\
$K$ & number of receivers in broadcast network\\
$L_n$ & duration of the $n$th block\\
%$\sigma^2$ & Background noise level\\
$h_n$ & channel in the $n$th block\\
$\beta$ & pathloss component of $h_n$\\
$z_n$ & fading component of $h_n$\\
%$\theta$ & Connection threshold for network topology\\
$p_n$ & transmit power used in the $n$th block\\
%{\color{blue}$\zeta_n$} & {\color{blue}SNR gap in the $n$th block}\\
$t$ & total number of bits of the message to transmit\\
$\epsilon$ & outage probability constraint\\
$\delta$ & latency constraint\\
$\pi_n$ & average delay of the $n$th feedback from the receiver\\
$\bar\pi$ & average delay of the feedback from the receiver\\
$r_n$ & achievable rate in block $n$\\
$c_n$ & mutual information in bits contained in block $n$\\
$C_n$ & accumulated mutual information after $n$ blocks\\
$f_n$ & probability density function (PDF) of $r_n$\\
$F_n$ & cumulative distribution function (CDF) of $r_n$\\
$g_n$ & PDF of $R_n$\\
$G_n$ & CDF of $R_n$\\
$Q_n$ & outage probability after $n$ blocks\\
%$\hat Q_n(t)$ & Estimated BLER based on network topology\\
$\hQ_n$ & proposed upper bound on $Q_n$\\
$\hhQ_n$ & existing upper bound on $Q_n$ in \cite{Laneman_Globecom03,Laneman_IT03}\\
%$f_X(x)$ & probability density function\\
%$F_X(x)$ & cumulative distribution function\\
%$\Delta$ & degrees of freedom in chi-square distribution\\
%$\mathcal U$ & set of remaining receivers\\
%$\mathcal U^c$ & set of nodes other than remaining receivers\\
\hline
%\hline
\end{tabular}}
\label{tab:notation}
\end{table}

\section{System Model}
\label{sec:sys}

We begin with a point-to-point channel of Rayleigh fading. Consider a sequence (in time) of channels $h_1,\ldots,h_N$ over $N$ blocks, each modeled as
\begin{equation}
\label{hn}
h_n = \sqrt{\beta}z_n,
\end{equation}
where the block index $n=1,\ldots,N$, the pathloss $\beta\in\mathbb R_+$ is fixed, and the Rayleigh fading component $z_n$ is drawn from the complex Gaussian distribution $\mathcal{CN}(0,1)$ independently across the blocks. Other types of fading, i.e., the Rician and the Nakagami, are discussed later in Section \ref{sec:multidim}.% We assume that only the pathloss $\beta$ is known \emph{a priori}.

The transmitter wishes to communicate a $t$-bit message to the receiver by using the $N$ blocks.
Let $p_n$ be the transmit power in the $n$th block and let $\sigma^2$ be the background noise power. Let $L_n$ be the duration of the $n$th block. With the bandwidth normalized to $1$ for ease of notation, the mutual information contained in the $n$th block can be computed as
\begin{align}
\label{r}
r_n &=\log_2\bigg(1+\frac{|h_n|^2p_n}{\sigma^2}\bigg),
\end{align}
so decoding the $n$th block \emph{alone} can recover at most $c_n$ bits, where
\begin{equation}
\label{c:1}
    c_n = L_nr_n.
\end{equation}
Define the pathloss-to-noise ratio to be
\begin{equation}
\label{s}
S = \frac{\beta}{\sigma^2}.
\end{equation}
Combining (\ref{hn})--(\ref{s}) together, we rewrite $c_n$ as
\begin{align}
\label{c:2}
c_n =L_n\log_2\left(1+S|z_n|^2 p_n \right).
\end{align}
The transmissions of these $N$ blocks are coordinated via HARQ as specified in the following subsection. In this paper, we assume that $L_n$'s are fixed and focus on the optimization of power $p_n$ across the $N$ transmissions.

\subsection{Chase Combining vs. Incremental Redundancy}
\label{subsec:HARQ}

The HARQ protocol works as follows. After each block $n=1,2,\ldots,N-1$, the receiver gives a feedback $\mathtt{ACK}/\mathtt{NACK}$ signal indicating whether the $t$-bit message has been successfully received or not, thereby either terminating HARQ or continuing to the next block $n+1$. HARQ finishes after the final block $N$ regardless of the message reception.

At the receiver side, the previous data packets are all saved and are \emph{jointly} decoded with the newly received packet in each block. Let $C_n$ be the total number of message bits that can be recovered by joint decoding after $n$ blocks. Notice that $C_n$ reduces to $c_n$ in (\ref{c:2}) if the receiver just discards all the old packets\footnote{This separate decoding scheme is referred to as type-I HARQ while the joint decoding scheme is referred to as type-II HARQ \cite{Vangelista_15,Berder_15}.}. There are two different variants of the joint decoding for HARQ:

\subsubsection{Chase Combining}

Each retransmission is identical to the initial transmission, so $L_n=L_1$ for any $n=2,\ldots,N$. The receiver treats each block as a diversity branch and performs the optimal coherent combination of them, thus achieving an energy accumulation:
\begin{equation}
\label{C:CC}
    C_n = L_1\log_2\left(1+S\sum^n_{i=1}|z_i|^2p_i\right).
\end{equation}

\subsubsection{Incremental Redundancy}
Alternatively, rather than repeating the original packet, the transmitter sends a different set of coded bits for the $t$-bit message in each retransmission, so the receiver obtains new redundant information about the message after each block. As shown in \cite{Caire_IT01}, the incremental redundancy coding can yield a mutual information accumulation:
\begin{subequations}
\label{C:IR}
\begin{align}
C_n &= \sum^n_{i=1}c_i\\
&=\sum^n_{i=1}L_i\log_2\left(1+S|z_i|^2 p_i \right).
\end{align}
\end{subequations}
%{\color{blue}We remark that the HARQ retransmissions (with incremental redundancy coding) can be assigned different redundancy versions as in 5G NR \cite{Dahlman_book18} so that $(L_1,\ldots,L_N)$ as well as $(S_1,\ldots,S)$ are distinct.}

From a power optimization perspective, incremental redundancy is more difficult to tackle than Chase combining because the power variables in (\ref{C:IR}) are contained in separate log terms\footnote{This structure leads us to successive convolutions when computing the distribution of $C_n$ as shown in Section \ref{subsec:actual_outage_prob}.}. The rest of the paper concentrates on incremental redundancy, except in Remark \ref{remark:CC}, Section \ref{sec:multidim} where the results are adapted to Chase combining.

%Incremental redundancy is more difficult to tackle than Chase combining in terms of the power optimization. The remainder of the paper focuses on incremental redundancy. In Section XXX, we show that our results can be readily adapted to Chase combining.

\subsection{Problem Formulation}

At block $n$, an outage occurs if $C_n$ is below the total number of bits $t$ that needs to be sent. We express this outage probability as
\begin{align}
\label{Qn:raw}
Q_n &= \Pr\big[C_n< t\big].
\end{align}
Because the entire HARQ procedure finishes after the final block $N$, the ultimate outage probability is given by $Q_N$.

Recall that block $n$ is transmitted if and only if the previous $n-1$ blocks are insufficient to convey the entire message, so the expected value of the total energy consumption throughout the HARQ procedure amounts to
\begin{subequations}
\label{E}
\begin{align}
E
&= \sum^N_{n=1}p_nL_n\Pr\big[C_{n-1}<t\big]\\
&= p_1L_1 + \sum^{N}_{n=2}p_{n}L_nQ_{n-1},
    \label{E:c}
\end{align}
\end{subequations}
where we let $C_0=0$ by convention; thus $\mathrm{Pr}[C_0<t]=1$. We use $\pi_n$ to denote the delay of the $n$th $\mathtt{ACK}/\mathtt{NACK}$ feedback from the receiver, $n=1,\ldots,N-1$. Assume that $\pi_n$'s are independent across the blocks, with the same expectation $\bar\pi$. 
The expected latency can be computed as
\begin{subequations}
\label{D}
\begin{align}
D 
&=L_1 + \mathbb{E}_{(\pi_1,\ldots,\pi_{N-1})}\left[\sum^{N}_{n=2} (L_n+\pi_{n-1})Q_{n-1}\right]\\
&= L_1 + \sum^{N}_{n=2} (L_n+\bar\pi)Q_{n-1}
\end{align}
\end{subequations}
The ultimate outage probability $Q_N$, the expected energy cost $E$, and the expected latency $D$ constitute the three key performance metrics, all of which are affected by the power variables $(p_1,\ldots,p_N)$.

We formulate a power control problem of minimizing $E$ while meeting the given requirements on $Q_N$ and $D$, along with a power constraint $P$ on each $p_n$. Let $\epsilon$ be the target outage probability and let $\delta$ be the target latency. The problem can be written as
\begin{subequations}
\label{prob}
\begin{align}
\underset{(p_1,\ldots,p_N)}{\text{minimize}} &\quad
  E
    \label{prob:obj}\\
{\text{subject to}}&\quad Q_N \le \epsilon,
    \label{prob:constr:1}\\
&\quad D \le \delta,\\
&\quad 0\le p_n\le P.
    \label{prob:constr:2}
%&\quad p_n\ge0,\;\forall n\in[1:N]\\
\end{align}
\end{subequations}
The full channel information of each $h_n$ is not available at the transmitter. Following \cite{Lee_TWC10,Larsson_TWC11,Su_TCOM11,Ge_IET15}, we assume that the transmitter only knows the distribution of $h_n$, i.e., the value of $\beta$ (hence $S$). Notice that all the power variables $(p_1,\ldots,p_N)$ must be decided before HARQ takes place, prior to any feedback from the receiver.

%we assume that the full channel state information (CSI) is not available in the above problem; only the pathloss $\beta$ and the distribution of the fading component $$
\setcounter{equation}{20}
\begin{figure*}
\begin{multline}
A_n(t) =
\frac{L_1S_1P(\ln2)^{n-1}}{\prod^n_{i=1}L_iS}\\
\cdot\bigg(\bigg[1-\exp\bigg(-\frac{2^{x_1/L_1}-1}{SP}\bigg)\bigg]*
\bigg[2^{x_2/L_2}\exp\bigg(-\frac{2^{x_2/L_2}-1}{SP}\bigg)\bigg]*\cdots*
\bigg[2^{x_n/L_n}\exp\bigg(-\frac{2^{x_{n}/L_n}-1}{SP}\bigg)\bigg]\bigg)(t)
    \label{An}
\end{multline}
\hrule
\end{figure*}
\setcounter{equation}{11}

The main obstacle for solving the power control problem in (\ref{prob}) is that none of $(Q_N,E,D)$ can be expressed in closed form in terms of the power variables. As shown in the next section, each $Q_n$ consists of successive convolutions that are computationally intractable. Since $E$ and $D$ both involve $Q_n$, the heart of the problem (\ref{prob}) is how to deal with the successive convolutions inside $Q_n$. This work overcomes the above obstacle by using a novel outage probability bound, which is tighter than the existing bound in \cite{Laneman_Globecom03,Laneman_IT03}, to isolate the power variables from the successive convolutions. With each $Q_n$ in (\ref{prob}) approximated by the new bound, we arrive at a GP problem that can be efficiently solved by standard optimization technique.

%Th yields an approximation of (\ref{prob}) in a GP form that is readily solvable by standard optimization techniques.

It is worth mentioning that alternative problem formulations involving the three performance metrics $(Q_N,E,D)$ are also possible, e.g., we could have minimized $D$ under the constraints on $E$ and $Q_N$. The proposed approach would work for these alternative problem formulations as well.

\section{A New Upper Bound on Outage Probability}
\label{sec:upper_bound}

\subsection{Actual Outage Probability}
\label{subsec:actual_outage_prob}

The last section only gives a raw expression of the outage probability $Q_n$ in (\ref{Qn:raw}). We need to further write $Q_n$ explicitly in terms of $(p_1,\ldots,p_N)$ in order to carry out the power optimization. Let us first compute the cumulative distribution function (CDF) of $c_n$ for each block $n$ as
\begin{subequations}
\begin{align}
F_{n}(x_n) &= \Pr\big[c_n<x_n\big]
    \label{Fn:a}\\
&=\Pr\left[L_n\log_2\left(1+S|z_n|^2p_n\right)<x_n\right]
    \label{Fn:a+}\\
&=\Pr\left[|z_n|^2<\frac{2^{x_n/L_n}-1}{Sp_n}\right]
    \label{Fn:b}\\
&=1-\exp\bigg(-\frac{2^{x_n/L_n}-1}{Sp_n}\bigg),
    \label{Fn:c}
\end{align}
\end{subequations}
where the last step follows by the Rayleigh distribution of the fading magnitude $|z_n|$. Further, the probability density function (PDF) of $c_n$ in (\ref{c:2}) can be obtained from $F_n(x_n)$ as
\begin{subequations}
\label{fn}
\begin{align}
f_{n}(x_n) &= \frac{d}{dx_n}F_{n}(x_n)
    \label{fn:a}\\
&= \frac{2^{x_n/L_n}\ln2}{L_nSp_n}\exp\bigg(-\frac{2^{x_n/L_n}-1}{Sp_n}\bigg).
    \label{fn:b}
\end{align}
\end{subequations}
We use $G_n(y_n)$ to denote the CDF of $C_n$ in (\ref{C:IR}), i.e.,
\begin{equation}
    G_n(y_n) = \mathrm{Pr}\left[C_n<y_n\right],
\end{equation}
and use $g_n(y_n)$ to denote the PDF of $C_n$.
Since $C_n=\sum^n_{i=1}c_i$, we can obtain the PDF of $C_n$ by computing the successive convolutions of the respective PDFs of $\{c_1,\ldots,c_n\}$ over the range $[0,y_n)$, i.e.,
\begin{subequations}
\begin{align}
\label{qn}
g_n(y_n) &= \idotsint\limits_\mathcal{
    \substack{
        0 \le x_i \le y_n,\,\forall i\\
        x_1+\cdots+x_n=y_n
    }
}f_1(x_1)\cdots f_n(x_n)dx_1\cdots dx_n\\
&= (f_1*f_2*\cdots*f_n)(y_n).
\end{align}
\end{subequations}
It follows that the CDF $G_n(y_n)$ can be recovered as
\begin{subequations}
\label{Gn}
\begin{align}
G_n(y_n) 
&= \int^{y_n}_0 g_n(\tau)d\tau
    \\
&= \bigg(\bigg(\int^{x_1}_0 f_1(\tau)d\tau\bigg)*f_2*\cdots*f_n\bigg)(y_n)
    \label{Gn:b}\\
&= (F_1*f_2*\cdots *f_n)(y_n),
    \label{Gn:c}
\end{align}
\end{subequations}
where (\ref{Gn:b}) follows by the property $\frac{d}{dx}\big(u(\tau)*v(\tau)\big)(x)=\big(\frac{d}{d\tau}u(\tau)*v(\tau)\big)(x)$. Recognize now the outage probability function $Q_n$ in (\ref{Qn:raw}) as the value of the CDF $G_n(y_n)$ with $y_n=t$, i.e.,
\begin{equation}
\label{Qn}
Q_n = G_n(t)=(F_1*f_2*\cdots *f_n)(t).
\end{equation}
We see that it is numerically difficult to optimize $(p_1,\ldots,p_N)$ in $Q_n$ directly because of the successive convolutions.

\subsection{Proposed Outage Probability Bound}

To make the power control problem (\ref{prob}) numerically tractable, we propose to approximate $Q_n$ in such a way that the power variables are isolated from the successive convolutions in (\ref{Qn}). Toward this end, we first relax each $f_n(x_n)$ in (\ref{fn}) by replacing $p_n$ with the highest possible transmit power $P$ in the exponential term, i.e.,
\begin{subequations}
\label{hf}
\begin{align}
f_n(x_n) &\le \hf_n(x_n)
    \label{hf:a}\\
&=
\frac{2^{x_n/L_n}\ln2}{L_nSp_n}\exp\bigg(-\frac{2^{x_n/L_n}-1}{SP}\bigg).
    \label{hf:b}
\end{align}
\end{subequations}
An upper bound on $F_n(x_n)$ immediately follows:
\begin{subequations}
\label{hF}
\begin{align}
F_n(x_n) &\le \hF_n(x_n)
    \label{hF:a}\\
&= \int^{x_n}_0\hf_n(\tau)d\tau
    \label{hF:b}\\
& = \frac{P}{p_n}\left(1-\exp\left(-\frac{2^{x_n/L_n}-1}{SP}\right)\right).
    \label{hF:c}
\end{align}
\end{subequations}
Furthermore, we substitute $\hf_n(x_n)$ and $\hF_n(x_n)$ into (\ref{Qn}) to replace $f_n(x_n)$ and $F_n(x_n)$, respectively, thereby obtaining an upper bound on $Q_n$ as
\begin{subequations}
\label{hQ}
\begin{align}
\hQ_n
&= (\hF_1*\hf_2*\cdots *\hf_n)(t)
    \label{hQ:b}\\
&= \frac{A_n(t)}{\prod^n_{i=1}p_i},
    \label{hQ:c}
\end{align}
\end{subequations}
where $A_n(t)$ is independent of the power variables as shown in (\ref{An}). The following proposition summarizes the above result.
\begin{proposition}[Outage Probability Bound for Rayleigh Fading]
\label{prop:Rayleigh}
For sending a $t$-bit message on i.i.d. Rayleigh fading channels, the outage probability $Q_n$ after $n$ blocks with incremental redundancy can be upper bounded as
\setcounter{equation}{21}
\begin{equation}
Q_n \le \hQ_n,
\end{equation}
where $\hat{Q}_n$ is given by (\ref{hQ}) and (\ref{An}). In particular, the equality holds if and only if $P\rightarrow\infty$.
\end{proposition}

We further show that the existing bound in \cite{Laneman_Globecom03,Laneman_IT03} is a special case of the proposed bound and is looser in general.

%The following proposition shows that the proposed upper bound $\hQ_n$ encompasses the existing upper bound in \cite{Laneman_Globecom03,Laneman_IT03} as a special case, .

\begin{proposition}[Connection to Existing Bound \cite{Laneman_Globecom03,Laneman_IT03}]
\label{prop:compare}
Assume that $L_n=L$ for any $n=1,\ldots,N$. As $P\rightarrow\infty$, the term $A_n(t)$ in (\ref{An}) reduces to
\begin{equation}
\label{A'}
A'_{n}(t) =\frac{(\ln2)^{n-1}}{L^{n-1}S^{n}}\\
\left(\big(2^{x_1/L}-1\big)*2^{x_2/L}*\cdots*2^{x_n/L}\right)(t)
\end{equation}
and accordingly the proposed upper bound $\hQ_n$ reduces to
\begin{equation}
\label{hhQ}
\hhQ_n = \frac{A'_{n}(t)}{\prod^n_{i=1}p_i},
\end{equation}
which is exactly the upper bound in \cite{Laneman_Globecom03,Laneman_IT03}. Moreover, for any message size $t>0$ and any block $n=1,2,\ldots,N$, we have
\begin{equation}
\label{bound_inequalities}
\hQ_n \le \hhQ_n,
\end{equation}
where the equality holds if and only if $P\rightarrow\infty$.
\end{proposition}
\begin{IEEEproof}
The reduction of $A_n(t)$ to $A'_n(t)$ can be verified by using the fact that $1-\exp\big(-\frac{2^{x_{1}/L}-1}{SP}\big)=\frac{2^{x_{1}/L}-1}{SP}$ and $\exp\big(-\frac{2^{x_{n}/L}-1}{SP}\big)=1$ as $P\rightarrow\infty$. Clearly, each $\hf_n(x_n)$ in (\ref{hf:b}) is a monotonically increasing function of $P$ for fixed $x_n$. With $\hQ_n$ in (\ref{hQ}) rewritten as $\hQ_n=\int^t_0 (\hf_1*\hf_2*\cdots*\hf_n)(\tau)d\tau$, we can show that $\hQ_n$ is also monotonically increasing with $P$. Since $\hhQ_n$ equals $\hQ_n$ as $P\rightarrow\infty$, we have $\hQ_n\le\hhQ_n$. The proof is then completed. %Notice that the proof would be much more cumbersome if we use the original form of $\hQ_n(t)$ in (\ref{hQ}) because $\hF_1(t)$ might not be an increasing function of $P$. 
%As $S\rightarrow\infty$, we have $A_n(t)=A_{n,\infty}(t)$ and $\hQ_n=\hhQ_n$; for $S<\infty$, we have $A_n(t)<A_{n,\infty}(t)$ and $\hQ_n<\hhQ_n$.
\end{IEEEproof}

\begin{figure}[t]
\begin{minipage}[b]{1.0\linewidth}
\psfrag{D}[][]{\footnotesize Dropout}
\psfrag{E}[][]{\footnotesize Phase $1$}
\psfrag{F}[][]{\footnotesize Phase $2$}
\centering
\vspace{-1em}
\centerline{\includegraphics[width=9.5cm]{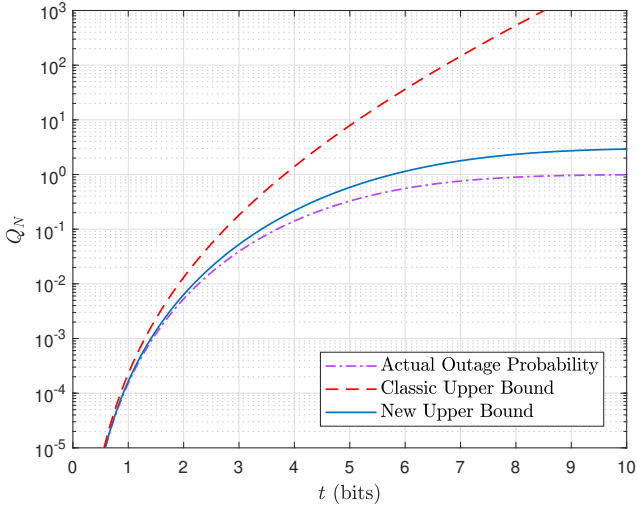}}
\caption{Actual outage probability vs. classic upper bound vs. proposed upper bound when $N=5$, $S=2$, $P=1$, and $p_n/P=0.8$ for any $n=1,\ldots,5$.}
\label{fig:bound}
\end{minipage}
\end{figure}

The original derivation of $\hhQ_n$ in \cite{Laneman_Globecom03,Laneman_IT03} is based on a \emph{piecewise squeezing} process, whereas we obtain the same result by specializing the proposed bounding technique in (\ref{hf})--(\ref{hhQ}). It turns out that the gap between the two bounds can be quite large, as shown in Fig.~\ref{fig:bound}.

\setcounter{equation}{38}
\begin{figure*}
\begin{align}
\label{check_An}
&{A}_n(t,\kappa) =\frac{L_1S_1P^\kappa(\kappa^\kappa\ln2)^{n-1}}{\Gamma^n(\kappa)\prod^n_{i=1}(L_iS)^\kappa}\Bigg(\gamma\left(\kappa,\frac{\kappa(2^{x_1/L_1}-1)}{S_1 P}\right)\notag\\
&\;\;*\left[(2^{x_2/L_2}-1)^{\kappa-1}2^{x_2/L_2}\exp\left(-\frac{\kappa(2^{x_2/L_n}-1)}{S_2P}\right)\right]*\cdots*
\left[(2^{x_n/L_n}-1)^{\kappa-1}2^{x_n/L_n}\exp\left(-\frac{\kappa(2^{x_n/L_n}-1)}{S P}\right)\right]\Bigg)(t)
\end{align}
\hrule
\end{figure*}

\section{Power Control via Geometric Programming}
\label{sec:link_GP}

Replacing each term $Q_n$ in (\ref{prob})  (including the ones contained in $E$ and $D$) with the proposed bound $\hQ_n$ gives rise to a GP problem:
\setcounter{equation}{25}
\begin{subequations}
\label{GP}
\begin{align}
\underset{(p_1,\ldots,p_N)}{\text{minimize}} &\quad
  p_1L_1+\sum^N_{n=2}\frac{L_np_nA_{n-1}(t)}{\prod^{n-1}_{i=1}p_i}
    \label{GP:obj}\\
{\text{subject to}}&\quad \frac{A_N(t)}{\prod^N_{i=1}p_i} \le \epsilon,
    \label{GP:constr:An}\\
&\quad \sum^{N-1}_{n=1}\frac{(L_{n+1}+\bar\pi)A_{n}(t)}{\prod^{n}_{i=1}p_i}\le \delta-L_1,
    \label{GP:constr:D}\\
&\quad 0\le p_n\le P.
    \label{GP:constr:P}
\end{align}
\end{subequations}
The optimal solution of the above problem can be efficiently obtained via the standard optimization technique. The following remarks and proposition provide some insights into this GP method and its solution.

\begin{remark}
    \label{remark:constraints}
It is crucial to approximate $Q_n$ from above. As a result, we would only overestimate the three performance metrics $(Q_N,E,D)$ so that the original constraints can be guaranteed.
\end{remark}

\begin{remark}
By approximating $Q_n$ as the monomial function $\hQ_n$, we basically approximate $E$ and $D$ as two posynomial functions. Thus, many other problem formulations of $(Q_N,E,D)$ can be cast into a GP form as well.
\end{remark}

\begin{remark}
\label{remark:tight_constraint}
Assuming that we do not have the power constraint (\ref{GP:constr:P}),
then the outage probability constraint (\ref{GP:constr:An}) must be tight at the optimum. This is because otherwise the objective function (\ref{GP:obj}) can be further decreased by scaling $p_N$ up to $A_N/\big(\epsilon \prod^{N-1}_{i=1}p_i\big)$; the new solution still satisfies the latency constraint since  (\ref{GP:constr:D}) does not include $p_N$. Unlike the outage probability constraint (\ref{GP:constr:An}), the latency constraint (\ref{GP:constr:An}) is not necessarily tight at the optimum.
\end{remark}

%We then examine a special case of (\ref{GP}) without the power constraint $P$.

\begin{proposition}
\label{prop:without_P}
Without the power constraint (\ref{GP:constr:D}), the optimal solution of the optimization problem would satisfy
\begin{equation}
\label{solu:2}
L_n\hQ_{n-1}p_n = 2L_{n+1}\hQ_np_{n+1}+\eta L_{n+1}\hQ_n,
\end{equation}
for any $n=1,\ldots,N-1$ and some $\eta\ge0$, where $\hat Q_{0}=1$ in particular.
\end{proposition}
\begin{IEEEproof}
First, move the latency constraint (\ref{GP:constr:D}) to the objective function in (\ref{GP:obj}) with a Lagrange multiplier $\eta\ge0$. Next, according to Remark \ref{remark:tight_constraint}, we substitute $p_N=A_N/\big(\epsilon \prod^{N-1}_{i=1}p_i\big)$ into the objective function. Further, we rewrite each power variable as $p_n = e^{y_n}$ so that the constraint $p_n\ge0$ can be automatically guaranteed. We then arrive at an unconstrained convex problem of $(y_1,\ldots,y_{N-1})$. Solving the first-order condition of the new problem yields
\begin{equation}
L_n e^{2y_n}A_{n-1}(t) = 2L_{n+1}e^{y_{n+1}}A_n(t)+\eta L_{n+1}A_n(t).
\end{equation}
We then obtain (\ref{solu:2}) by substituting $y_n=\ln p_n$ into the above equation.
\end{IEEEproof}

%The above proposition further provides a new insight into the problem setting of URLLC as discussed in what follows.

\begin{remark}%[Necessity of power constraint for ultrareliable communications]
\label{remark:unbouned}
One might postulate that minimizing $E$ alone suffices to suppress the transmit power. However, this is not the case. According to Proposition \ref{prop:without_P}, the solution of the GP problem (\ref{GP}) without the latency constraint in (\ref{GP:constr:D}) and the maximum power constraint in (\ref{GP:constr:P}) satisfies 
$L_n\hQ_{n-1}p_n = 2L_{n+1}\hQ_np_{n+1}$, for each $n=1,\ldots,N-1$. Combining the above $N-1$ equations together, we have
\begin{equation}
\frac{p_N}{p_1} = \frac{L_1}{2^{N-1}\hQ_{N-1}L_N}.
\end{equation}
As a consequence, $p_N$ can be much higher than $p_1$. For instance, the approximated outage probability $\hat  Q_{N-1}$ typically lies in the interval $[10^{-9},10^{-5}]$ in URLLC, and typically $N=4$ in 5G NR \cite{Dahlman_book18}. Thus, without the max power constraint, the transmit power of the final packet can be approximately $41-81$dB higher than that of the initial packet. Thus, minimizing $E$ alone can only suppress the expected energy cost, but the individual transmit power may still spike. For this reason, we conclude that it is necessary to impose a per block power constraint $P$ in practice. The above analysis is demonstrated numerically in Fig.~\ref{fig:pn_1AT_unbounded} in Section \ref{sec:simulations}.
\end{remark}

\begin{remark}
The earlier works \cite{Eriksson_14,Szczecinski_TCOM16}
show that the optimal power satisfies $L_1p_1<L_2p_2<\ldots<L_Np_N$, i.e., the energy increases in each round, when minimizing the expected energy consumption with the target outage probability. But this is not the case in our problem because of the additional constraints on latency and transmit power.
\end{remark}

\section{Generalized Outage Probability Bound}
\label{sec:multidim}

We now extend the proposed bound and GP method to more general wireless environment. The extensions are threefold. First, we consider a parametric Nakagami fading that can imitate a variety of channels (including Rician fading and  Rayleigh fading). Second, we discuss the multi-antenna channels, and show that HARQ with Chase combining can be addressed similarly. Third, message broadcast to multiple receivers is studied.
%The goal of this section is to generalize the proposed bound $\hQ_n(t)$ and GP method to a broader range of channel models. We begin with the Nakagami fading---a flexible parameterized model that is capable of mimicing the Rician fading as well as a variety of practical fading. The second part of the section focuses on the Rayleigh fading and introduces spatial diversity by assuming multiple transmit antennas.%; the extension 

%that either transmitter or receiver has multiple antennas; in statistics terms, the extension aims to account 

%In statistics terms, we now consider the chi-square distribution and the Nakagami distribution that encompass the Rayleigh distribution as a special case. In wireless communication terms, the generalized GP method is capable of reaping the antenna diversiy gain.

%This section generalizes the proposed upper bound $\hQ_n$ to account for vector channels, along with the GP-based power control method extended accordingly.
\setcounter{equation}{49}
\begin{figure*}
\begin{multline}
\label{sys:An}
{A}_n(t,M) = \frac{L_1S_1^MP^M(\ln2)^{n-1}}{((M-1)!)^nS^{Mn}\prod^n_{i=1}L_i}\Bigg(\gamma\left(M,\frac{2^{x_1/L_1}-1}{S_1 P}\right)\\
\qquad\;*
\bigg[(2^{x_2/L_2}-1)^{M-1}2^{{x_2/L_2}}\exp\left(-\frac{2^{x_2/L_2}-1}{S_2 P}\right)\bigg]*\cdots*
\bigg[(2^{x_n/L_n}-1)^{M-1}2^{x_n/L_n}\exp\left(-\frac{2^{x_n/L_n}-1}{S P}\right)\bigg]\Bigg)(t)
\end{multline}
\hrule
\end{figure*}
\setcounter{equation}{29}

\subsection{Nakagami Fading}
\label{subsec:Nakagami}

In this subsection we still adopt the system model as described in Section \ref{sec:sys}, but shift attention from Rayleigh fading to Nakagami fading. Specifically, we assume that the fading magnitude $|z_n|$ has a Nakagami distribution with the parameter $\kappa>\frac12$ \cite{Goldsmith_book05}. The corresponding PDF of the fading power
\setcounter{equation}{29}
\begin{equation}
\label{lambda}
    \lambda_n = |z_n|^2
\end{equation}
is then given by
\begin{equation}
    \Lambda_n(\lambda_n) = \frac{\lambda_n^{\kappa-1}\kappa^\kappa}{\Gamma(\kappa)}\exp(-\kappa\lambda_n),
\end{equation}
where $\Gamma(\kappa)$ refers to the Gamma function
\begin{equation}
\label{Gamma}
\Gamma(\kappa) = \int^\infty_0\tau^{\kappa-1}\exp(-\tau)d\tau.
\end{equation}
As shown in \cite{Goldsmith_book05}, Nakagami fading is a general channel model that can be parameterized by $\kappa$ to fit a variety of empirical measurements.
Plugging (\ref{lambda}) into (\ref{c:2}), we can treat $\lambda_n$ as a function of $c_n$, i.e.,
\begin{equation}
    \label{Nakagami:rn}
    \lambda_n(c_n) = \frac{2^{c_n/L_n}-1}{Sp_n},
\end{equation}
Because $\lambda_n(c_n)$ is a monotonic function, the PDF of $c_n$ can be obtained from the PDF of $\lambda_n$ by applying the change of variable theorem, i.e.,
\begin{subequations}
\begin{align}
    f_{n}(x_n)
    &= \Lambda_n\left(\lambda_n(x_n)\right)\cdot\frac{d\lambda_n(x_n)}{dx_n}\\
    &= \frac{\kappa^\kappa(2^{x_n/L_n}-1)^{\kappa-1}2^{x_n/L_n}\ln2}{L_n\Gamma(\kappa)(Sp_n)^\kappa}\notag\\
    &\qquad\qquad\qquad\quad\cdot\exp\bigg(-\frac{\kappa(2^{x_n/L_n}-1)}{Sp_n}\bigg).
\end{align}
\end{subequations}
Following the bounding technique stated in Section \ref{sec:upper_bound} for Rayleigh fading, we first relax the exponential part of $f_n(x_n)$ to find an upper bound on $f_n(x_n)$:
\begin{multline}
    \label{Nakagami:hf}
    \hat f_{n}(x_n) = \frac{\kappa^\kappa(2^{x_n/L_n}-1)^{\kappa-1}2^{x_n/L_n}\ln2}{L_n\Gamma(\kappa)(Sp_n)^\kappa}\\\cdot\exp\bigg(-\frac{\kappa(2^{x_n/L_n}-1)}{SP}\bigg).
\end{multline}
Next, taking the integral of $\hat f_n(x_n)$ leads us to an upper bound on the CDF $F_n(x_n)$:
\begin{equation}
    \label{Nakagami:hF}
    \hat F_{n}(x_n) = \frac{1}{\Gamma(\kappa)}\bigg(\frac{P}{p_n}\bigg)^\kappa\cdot\gamma\bigg(\kappa,\frac{\kappa(2^{x_n/L_n}-1)}{SP}\bigg),
\end{equation}
where $\Gamma(\cdot)$ is the Gamma function in (\ref{Gamma}) and $\gamma(\cdot,\cdot)$ is the {lower incomplete Gamma function}
\begin{equation}
\gamma(a,b) = \int^b_0\tau^{a-1}\exp(-\tau)d\tau.
\end{equation}
% (which equals to $m!$ when $m\in\mathbb N$).
Substituting (\ref{Nakagami:hf}) and (\ref{Nakagami:hF}) back into (\ref{hQ}) yields the following upper bound on $Q_n$:
\begin{align}
\label{Nakagami:hQ}
\hQ_n &= \frac{A_n(t,\kappa)}{\prod^n_{i=1}p_i^{\kappa}},
\end{align}
where $A_n(t,\kappa)$ is computed as in (\ref{check_An}). We formalize the above extension in the following proposition.

\begin{proposition}[Outage Probability Bound for Nakagami Fading]
\label{prop:bound:Nakagami}
For sending a $t$-bit message on i.i.d. Nakagami fading channel, the outage probability $Q_n$ after $n$ blocks with incremental redundancy can be upper bounded as
\setcounter{equation}{39}
\begin{equation}
    \label{bound_inequality:Nakagami}
Q_n \le \hQ_n,
\end{equation}
where $\hat{Q}_n$ is given by (\ref{Nakagami:hQ}) and (\ref{check_An}). In particular, the equality holds if and only if $P\rightarrow\infty$.
\end{proposition}

Because the outage probability bound $\hQ_n$ in (\ref{Nakagami:hQ}) remains a monomial, the GP method proposed in Section \ref{sec:link_GP} for Rayleigh fading continues to work for Nakagami fading. Most importantly, we can now address a broad class of fading models by adjusting the Nakagami parameter $\kappa$, as illustrated in the following remarks.

\begin{remark}[Specialization to Rayleigh Fading]
\label{}
The outage probability bound of Nakagami fading in Proposition \ref{prop:bound:Nakagami} reduces to that of Rayleigh fading in Proposition \ref{prop:Rayleigh} when $\kappa=1$.
\end{remark}

\begin{remark}[Specialization to Rician Fading]
\label{}
The outage probability bound of Nakagami fading in Proposition \ref{prop:bound:Nakagami} account for Rician fading with the parameter $\mu$ by letting $\kappa=(\mu+1)^2/(2\mu+1)$ \cite{Goldsmith_book05}.
\end{remark}

%In addition, because the Nakagami distribution can be used to approximate the Rician distribution \cite{Goldsmith_book05}, Proposition \ref{prop:bound:Nakagami} can be correspondingly carried over to the Rician fading as stated in the following corollary.
%\begin{corollary}[Upper Bound on Outage Probability in Rician Fading Channel]
%\label{prop:bound:Rician}
%For the same setting as in Proposition \ref{prop:bound} except that the fading envelope $|g_{n}|$ is modeled as an i.i.d. Rician random variable with parameter $\mu$, $\hQ_n$ given by (\ref{Nakagami:hQ}) and (\ref{check_An}) with $m=(\mu+1)^2/(2\mu+1)$ is approximately an upper bound on the outage probability $Q_n$ after $n$ transmissions.
%\end{corollary}

\subsection{Multi-Antenna Channels}
\label{subsec:multi_antenna}

We now consider a multiple-input single-output (MISO) channel with Rayleigh fading. In statistics terms, it entails extending the new bound to higher degree-of-freedom of the chi-square distribution. Because the diversity gain by multi-antenna transmission resembles the energy accumulation effect, the extended bounding technique applies to HARQ with Chase combining as well.

%In statistics terms, the extended new bound proposed in this section can be thought of as an extension of the chi-square distribution from 2 degrees of freedom to higher degrees of freedom; in wireless communication terms, the generalized upper bound takes advantage of the spatial diversity provided by multiple antennas.

Assume that the transmitter has $M>1$ antennas. Use $h_{jn}\in\mathbb C$ to denote the channel from the $j$th transmit antenna to the receiver in block $n$. As before, each channel $h_{jn}$ is decomposed into two parts:
\begin{equation}
h_{jn} = \sqrt{\beta}z_{jn}.
\end{equation}
We assume that all these $h_{jn}$'s have a common fixed pathloss component $\beta\in\mathbb R_+$ and i.i.d. fading components $z_{jn}$'s drawn from $\mathcal{CN}(0,1)$. Recall that the channel information is unknown at the transmitter, so uniform power allocation across the $M$ transmit antennas is the best strategy \cite{Goldsmith_book05}. As a result, the number of message bits obtained from block $n$ alone without incremental redundancy can be computed as
\begin{equation}
\label{MIMO:cn}
c_{n} 
=L_n\log_2\Bigg(1+S\sum^M_{j=1}|z_{jn}|^2p_n\Bigg).
\end{equation}
Define $\nu_n\in\mathbb R_+$ to be two times the sum of the squares of $|z_{jn}|$, $j=1,\ldots,M$, i.e.,
\begin{equation}
\label{nu}
    \nu_n = 2\sum^M_{j=1}|z_{jn}|^2.
\end{equation}
Because each $z_{jn}\sim\mathcal{CN}(0,1)$, $\nu_n$ can be recognized as the sum of the squares of $2M$ independent standard real Gaussian random variables, i.e., $\nu_n$  has a chi-square distribution with $2M$ degrees of freedom. Substituting $\nu_n$ in (\ref{nu}), we can rewrite $c_n$ in (\ref{MIMO:cn}) as
\begin{align}
\label{multi:Rayleigh:rn}
c_{n}&=\log_2\left(1+\frac{1}{2}Sp_n\nu_n\right).
\end{align}
We remark that the above $c_n$ is also achievable for a single-input multiple-output (SIMO) channel with $M$ receive antennas. In that case, the receiver achieves $c_n$ in (\ref{multi:Rayleigh:rn}) by means of maximum ratio combining (MRC). %Furthermore, these results can be extended to the 

For the multi-antenna version of $c_n$ in (\ref{multi:Rayleigh:rn}) with diversity gain, its CDF can be computed as
\begin{subequations}
\label{sys:Fn:1}
\begin{align}
{F}_{n}(x_n)
&=\Pr\left[c_n<x_n\right]\\
&= \Pr\left[\nu_n<\frac{2(2^{x_n/L_n}-1)}{S p_n}\right]\\
&= \frac{1}{(M-1)!}\cdot\gamma\left(M,\frac{2^{x_n/L_n}-1}{S p_n}\right).
\end{align}
\end{subequations}
Taking the derivative of $F_n(x_n)$ gives the PDF of $c_n$, i.e.,
\begin{multline}
f_{n}(x_n)
=\frac{(2^{x_n/L_n}-1)^{M-1}2^{x_n/L_n}\ln2}{L_n(M-1)!(Sp_n)^{M}}\\
\cdot\exp
\left(-\frac{2^{x_n/L_n}-1}{S p_n}\right).
\end{multline}
Again, we use the power constraint $P$ to relax the exponential term of $f_n(x_n)$ and thereby construct an upper bound
\begin{multline}
    \hf_{n}(x_n)
    =\frac{(2^{x_n/L_n}-1)^{M-1}2^{x_n/L_n}\ln2}{L_n(M-1)!(Sp_n)^{M}}\\
    \cdot\exp
    \left(-\frac{2^{x_n/L_n}-1}{S P}\right).
    \end{multline}
The integral of $\hf_n(x_n)$ gives an upper bound on $F_n(x_n)$:
\begin{equation}
\label{MIMO:hF}
\hF_{n}(x_n)
= \frac{1}{(M-1)!}\bigg(\frac{P}{p_n}\bigg)^{M}\gamma\left(M,\frac{2^{x_n/L_n}-1}{S P}\right).
\end{equation}
Furthermore, substituting the above $\hf_{n}(x_n)$ and $\hF_{n}(x_n)$ into (\ref{hQ:b}) gives an upper bound on the outage probability $Q_n$ as
\begin{align}
\label{multi:hQ}
\hQ_n &= \frac{{A}_n(t,M)}{\prod^n_{i=1}p_i^{M}},
\end{align}
where $A_n(t,M)$ is given by (\ref{sys:An}). The following proposition summarizes the multi-antenna extension.

%The above upper bound is stated in the following proposition.
\begin{proposition}[Outage Probability Bound for multi-antenna Rayleigh Fading]
\label{prop:bound:multiantenna}
For sending a $t$-bit message on i.i.d. $M\times1$ MISO (or $1\times M$ SIMO) Rayleigh fading channel, the outage probability $Q_n$ after $n$ blocks with incremental redundancy can be upper bounded as
\setcounter{equation}{50}
\begin{equation}
Q_n \le \hQ_n,
\end{equation}
where $\hat{Q}_n$ is given by (\ref{sys:An}) and (\ref{multi:hQ}). In particular, the equality holds if and only if $P\rightarrow\infty$.
\end{proposition}

\begin{proposition}[Connection to Existing Bound in \cite{Ge_IET15}]
\label{prop:compare:multi}
Assume that $L_n=L$ for any $n=1\ldots,N$. As $P\rightarrow\infty$, the term $A_n(t,M)$ in (\ref{sys:An}) reduces to
\begin{align}
\label{MIMO:A'}
&A'_{n}(t,M) = \frac{(\ln2)^{n-1}}{M((M-1)!)^{n}S^{nM}}\Big((2^{x_1/L}-1)^M\,*\notag
\\
&\Big[2^{x_2/L}\big(2^{x_2/L}-1)^{M-1}\Big]*\cdots*\Big[2^{x_n/L}(2^{x_/Ln}-1)^{M-1}\Big]\Big)(t)
\end{align}
and accordingly the proposed upper bound $\hQ_n$ reduces to
\begin{equation}
\label{multi:hhQ}
\hhQ_n = \frac{A'_{n}(t,M)}{\prod^n_{i=1}p_i^{M}},
\end{equation}
which is the upper bound in \cite{Ge_IET15}. Moreover, for any message size $t>0$ and any block $n=1,2,\ldots,N$, we have
\begin{equation}
\label{bound_inequalities:multi}
\hQ_n \le \hhQ_n,
\end{equation}
where the equality holds if and only if $P\rightarrow\infty$.
\end{proposition}
\begin{IEEEproof}
The reduction of $A_n(t,M)$ in (\ref{sys:An}) to $A'_n(t,M)$ in (\ref{MIMO:A'}) relies on the property of the lower incomplete gamma function that $\gamma(a,b)/b^a=1/a$ as $b\rightarrow0$. The rest of the proof follows that of Proposition \ref{prop:compare} closely.
\end{IEEEproof}

Using the extended bound $\hQ_n$ to approximate $Q_n$ throughout the problem (\ref{prob}), we again obtain a GP problem of the power variables which can be solved efficiently. Moreover, several observations in the above extension are worth noting.

\begin{remark}[MIMO Channels]
For a general MIMO channel with stream multiplexing, the value of $c_n$ depends on the singular values of each channel realization, the distribution of which is computationally intractable. Nevertheless, if we only utilize MIMO to boost diversity, then the power control problem is structured as in the MISO case and can be readily addressed by the above approach.
\end{remark}

\begin{remark}[Chase combining]
\label{remark:CC}
Notice that the $c_n$ expression (\ref{MIMO:cn}) resembles the $C_n$ expression (\ref{C:CC}) for Chase combining when $p_n=p, n=1,\ldots,N$. Thus, we can readily obtain the following outage probability bound by applying the bounding technique in (\ref{MIMO:hF}) to $C_n$:
\begin{equation}
    \hQ_n=\frac{1}{(n-1)!}\bigg(\frac{P}{p}\bigg)^{n}\gamma\left(n,\frac{2^{t/L}-1}{S P}\right).
\end{equation}
Since $\hQ_n$ is still a monomial function of $p_n$, the proposed GP method also works for HARQ with Chase combining.
\end{remark}

%It is worthwhile to compare

\subsection{Broadcast Channels}

As a further extension, consider a broadcast channel in which the transmitter wishes to deliver a common $t$-bit message to $K\ge2$ receivers. Assume that the channel fadings are independent among the receivers.
We remark that the channel model can vary from receiver to receiver in terms of the fading distribution and the number of receive antennas. Let $C_{kn}$ be the accumulated number of message bits recovered by the $k$th receiver in block $n$. For each particular receiver $k$ in block $n$, we define its outage probability as $Q_{kn}=\Pr[C_{kn}<t]$, for which an upper bound $\hQ_{kn}$ can be readily constructed by using the previous technique. 

We still consider the power control problem in (\ref{prob}).
An outage occurs in block $n$ if at least one receiver $k$ fails to recover the entire message at this point. The overall outage probability in block $n$ is defined and bounded as follows:
\begin{subequations}
\begin{align}
    Q_n &= \mathrm{Pr}\left[C_{kn}<t \;\text{for at least one}\;k\right]\\
    &\le \sum^K_{k=1} \mathrm{Pr}\left[C_{kn}<t\right]
        \label{BC:bound:b}\\
    &\le \sum^K_{k=1} \hQ_{kn},
        \label{BC:bound:c}
\end{align}
\end{subequations}
where (\ref{BC:bound:b}) follows by the union bound while (\ref{BC:bound:c}) follows by the individual outage probability bound. Recall that each $\hQ_{kn}$ is a monomial function of $(p_1,\ldots,p_n)$ regardless of the channel model associated with receiver $k$, so the overall outage probability bound in (\ref{BC:bound:c}) is a posynomial function of $(p_1,\ldots,p_n)$. With each $Q_n$ approximated by the overall outage probability bound in (\ref{BC:bound:c}), we again convert the original problem (\ref{prob}) to a GP problem.

Furthermore, notice that the transmitter only needs to know whether there exist any remaining receivers which have not successfully decoded the message after each block. Since the transmitter does not need to identify the receivers, we can just let the remaining receivers send back their 1-bit $\mathtt{NACK}$ signals over the air without multiplexing, thereby significantly reducing the cost of user detection in massive IoT. %We also remark that modulation and coding scheme (MCS) of low order (e.g., BPSK) and energy efficient waveform (e.g., filter-bank multi-carrier) are preferred for the NACK signal sending back because of the low rate transmission and the transmitter hardware restriction in this case.}

\section{Simulation Results}
\label{sec:simulations}

This section numerically demonstrates the advantage of the proposed new outage probability bound and the resulting GP-based power control method as compared to the existing bound in \cite{Laneman_Globecom03,Laneman_IT03}. The  bandwidth is normalized to 1. The maximum power $P$ is also normalized to 1. Assume that the blocks are of the same length $L$; we measure latency in terms of $L$. Let $E_0$ be the energy consumption of transmitting a single block at full power; we measure energy cost in terms of $E_0$. The simulation setting follows: 
the total number of blocks $N=5$, the average latency of feedback $\bar\pi=0$, the message size $t=4$ bits, the outage probability constraint $\epsilon=10^{-5}$, and the latency constraint $\delta=3$ units. 
Throughout the section, we consider HARQ with incremental redundancy and assume Rayleigh fading. The value of $C_n$ is computed as in (\ref{C:IR}). For the broadcast channel, we assume that the multiple receivers have the same pathloss-to-noise ratio $S$.
The energy cost $E$ and the latency $D$ are both empirically evaluated by averaging out a large number of random trials.

\begin{figure}[t]
\begin{minipage}[b]{1.0\linewidth}
\centering
\centerline{\includegraphics[width=9.5cm]{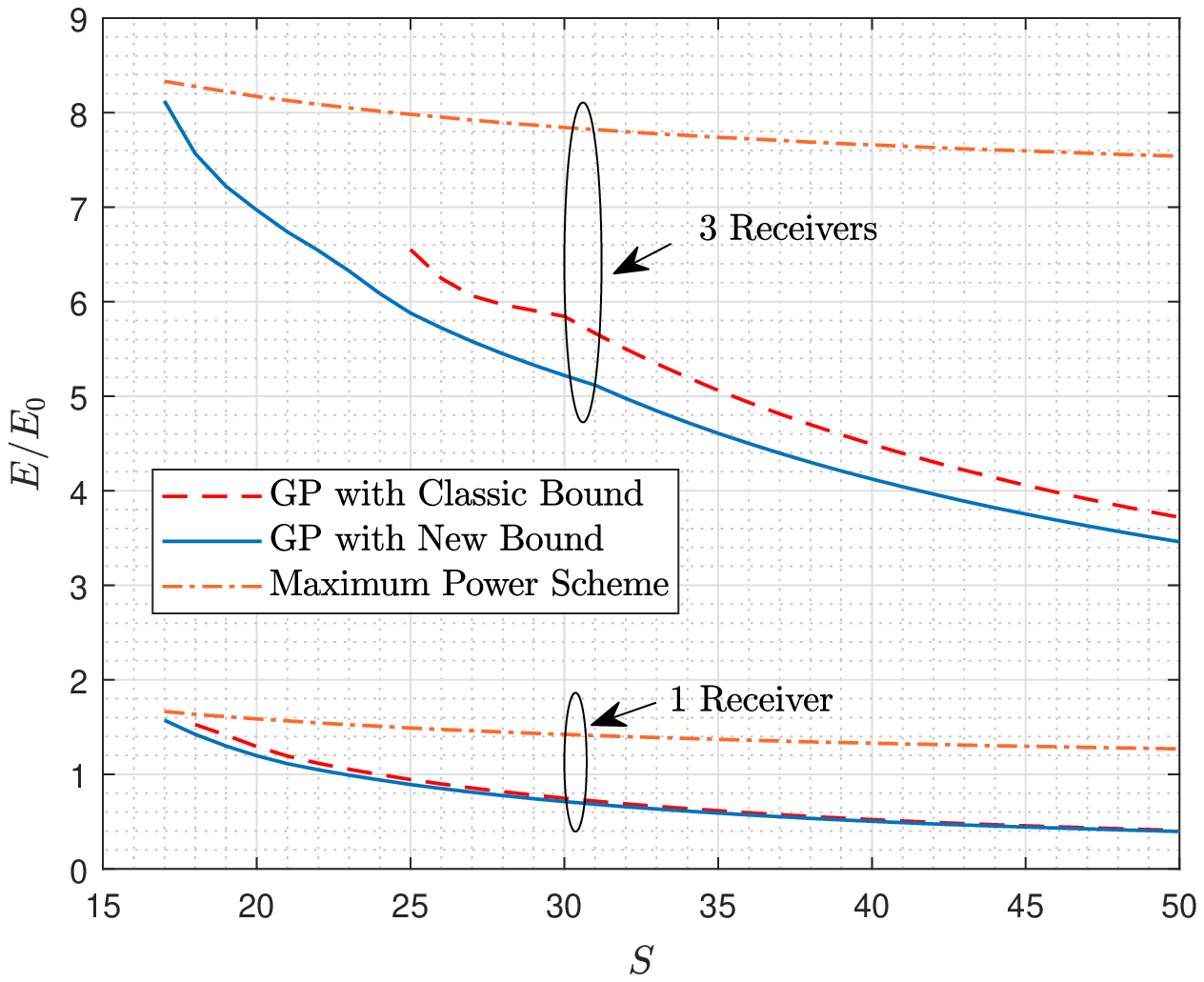}}
\vspace{-1em}
\caption{Energy vs. pathloss-to-noise ratio when each receiver has $1$ antenna.}
\label{fig:ES_1AT}
%\vspace{1em}
\end{minipage}
\begin{minipage}[b]{1.0\linewidth}
\centering
\centerline{\includegraphics[width=9.5cm]{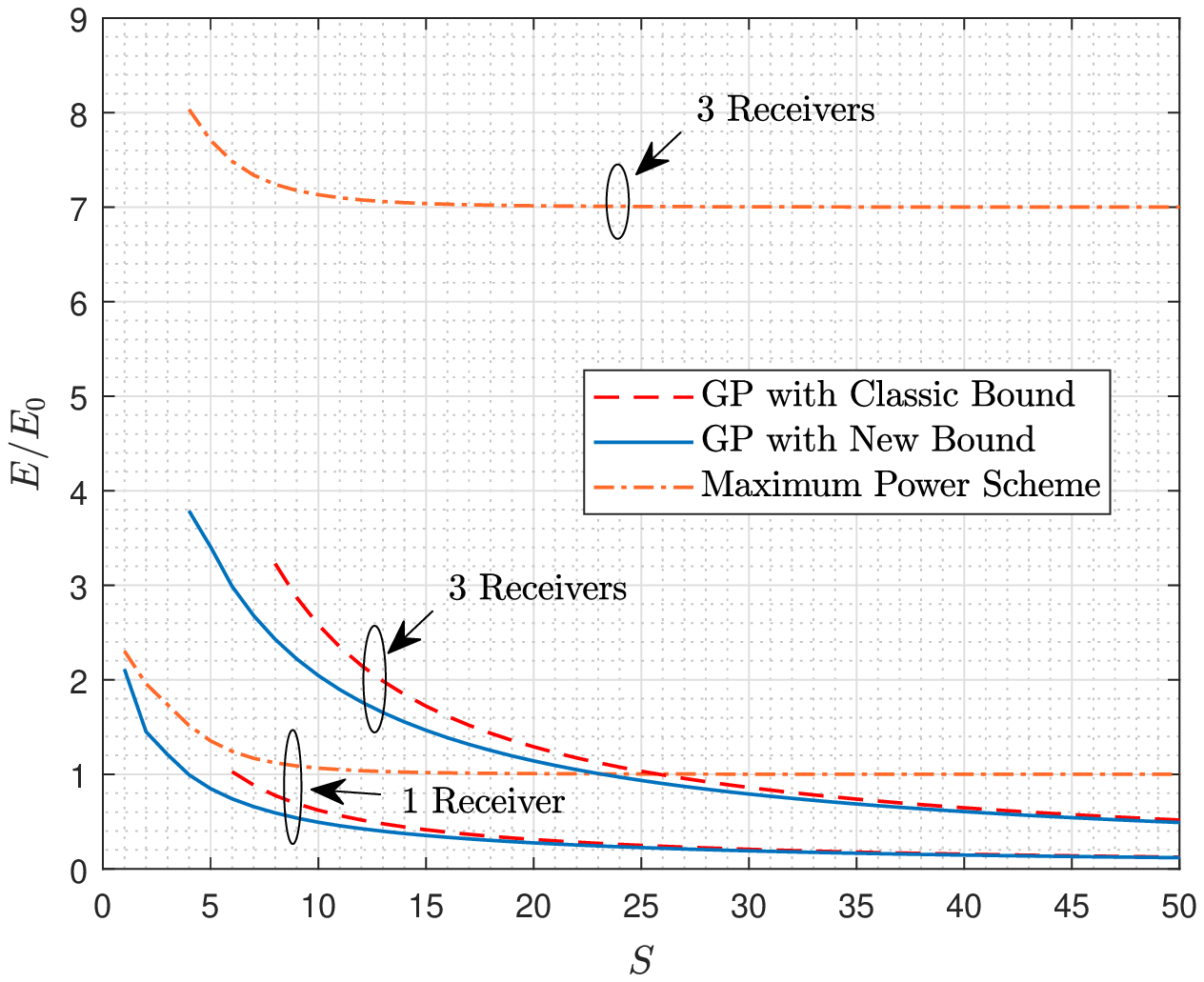}}
\vspace{-1em}
\caption{Energy vs. pathloss-to-noise ratio when each receiver has $4$ antennas.}
\label{fig:ES_4AT}
%\vspace{1em}
\end{minipage}
\end{figure}

Fig.~\ref{fig:ES_1AT} shows the energy cost versus the pathloss-to-noise ratio $S$ when the transmitter and receiver(s) have one antenna each. We consider both the point-to-point channel and the three-receiver broadcast channel in the figure. Observe that the two GP methods both reduce the energy cost significantly as compared to the maximum power scheme. For instance, when $S=50$, the energy cost is decreased by more than 50\% for the broadcast channel and is decreased by almost 70\% for the point-to-point channel. 
Observe also that the two GP methods have close performance when applied to the point-to-point channel. Nevertheless, in the presence of three receivers, GP with the new bound becomes much superior to GP with classic bound. Because the actual outage probability is considerably overestimated by the classic bound, the corresponding GP method wrongly suggests that the target outage probability $10^{-5}$ cannot be achieved unless $S$ is higher than 25. In comparison, GP with the new bound can work with much lower $S$; this is an admirable quality of the new bound for the low-power IoT applications. Further, Fig.~\ref{fig:ES_4AT} shows the SIMO case in which each receiver has 4 antennas. It can be seen that the gap between the maximum power scheme and the GP methods becomes larger when more antennas are deployed. The new bound still enables the GP method to work in a broader range of $S$ as compared to the classic bound.

%============================
\begin{figure}[t]
\begin{minipage}[b]{1.0\linewidth}
\centering
\centerline{\includegraphics[width=9.5cm]{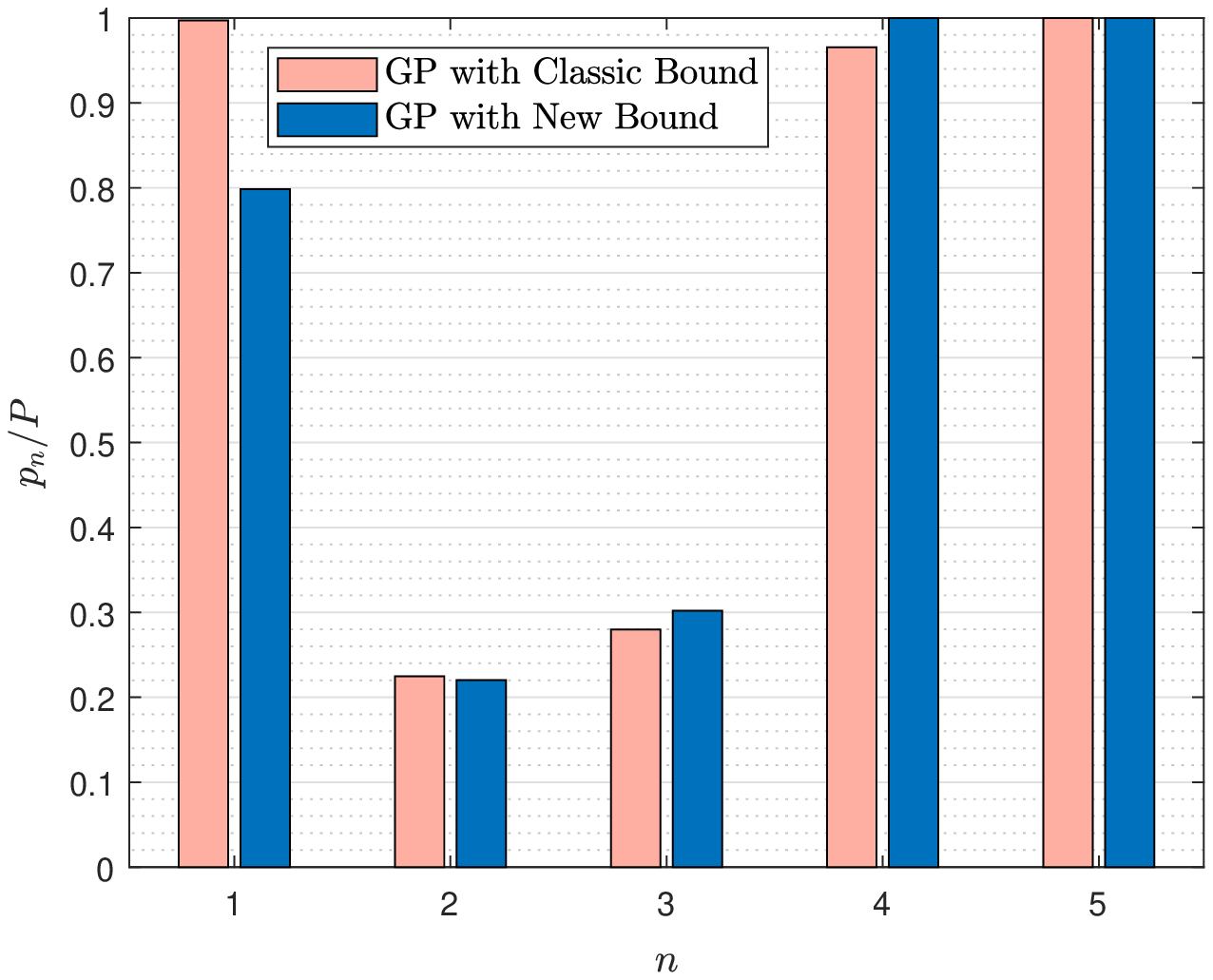}}
\vspace{-1em}
\caption{Transmit power in each round of HARQ for a SISO channel.}
\label{fig:pn_1AT}
%\vspace{1em}
\end{minipage}
\begin{minipage}[b]{1.0\linewidth}
\centering
\centerline{\includegraphics[width=9.5cm]{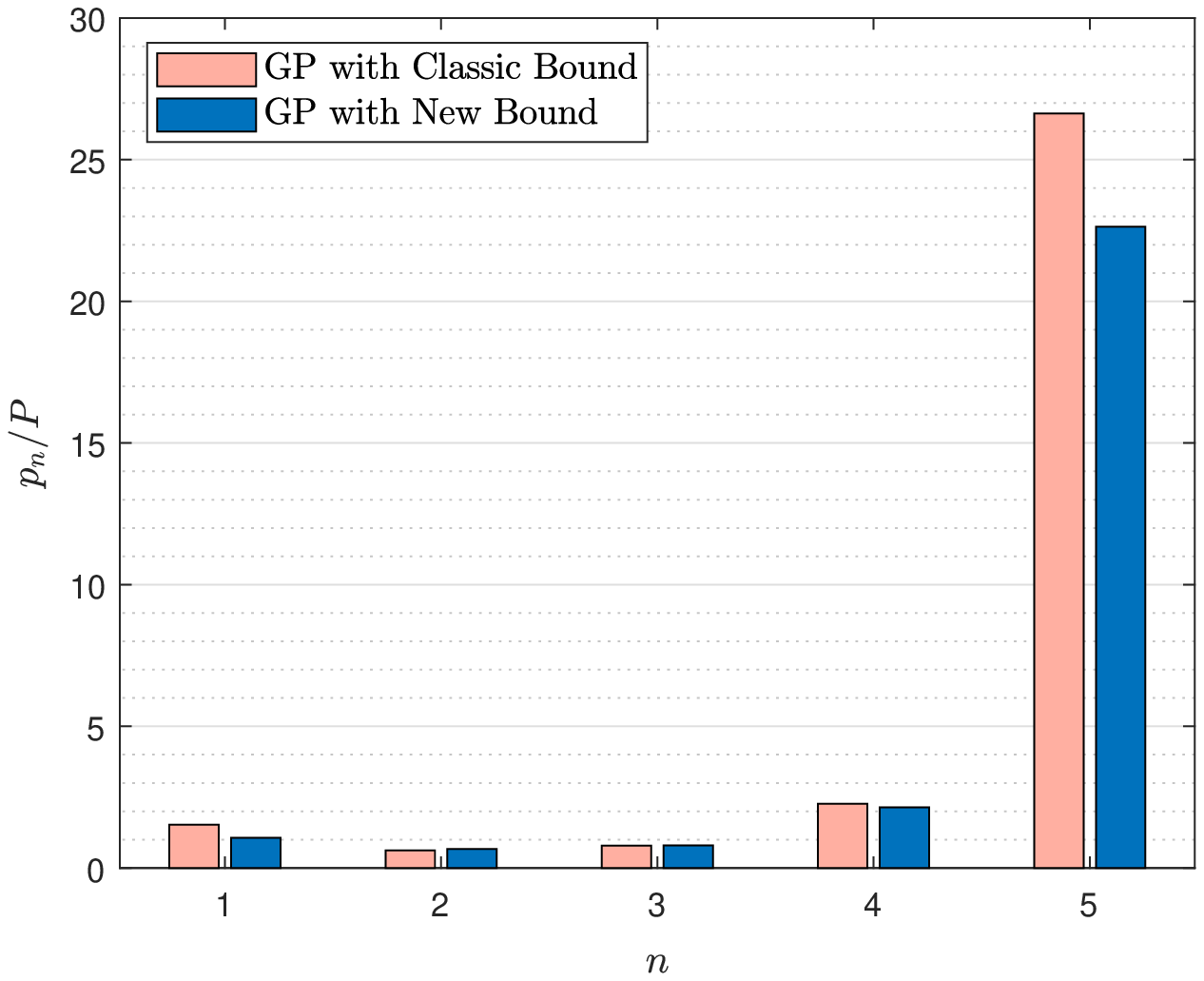}}
\vspace{-1em}
\caption{Transmit power in each round of HARQ for a SISO channel without the max power constraint and the latency constraint.}
\label{fig:pn_1AT_unbounded}
\vspace{-0.4em}
\end{minipage}
\end{figure}

\begin{figure}
\begin{minipage}[b]{1.0\linewidth}
\centering
\centerline{\includegraphics[width=9.5cm]{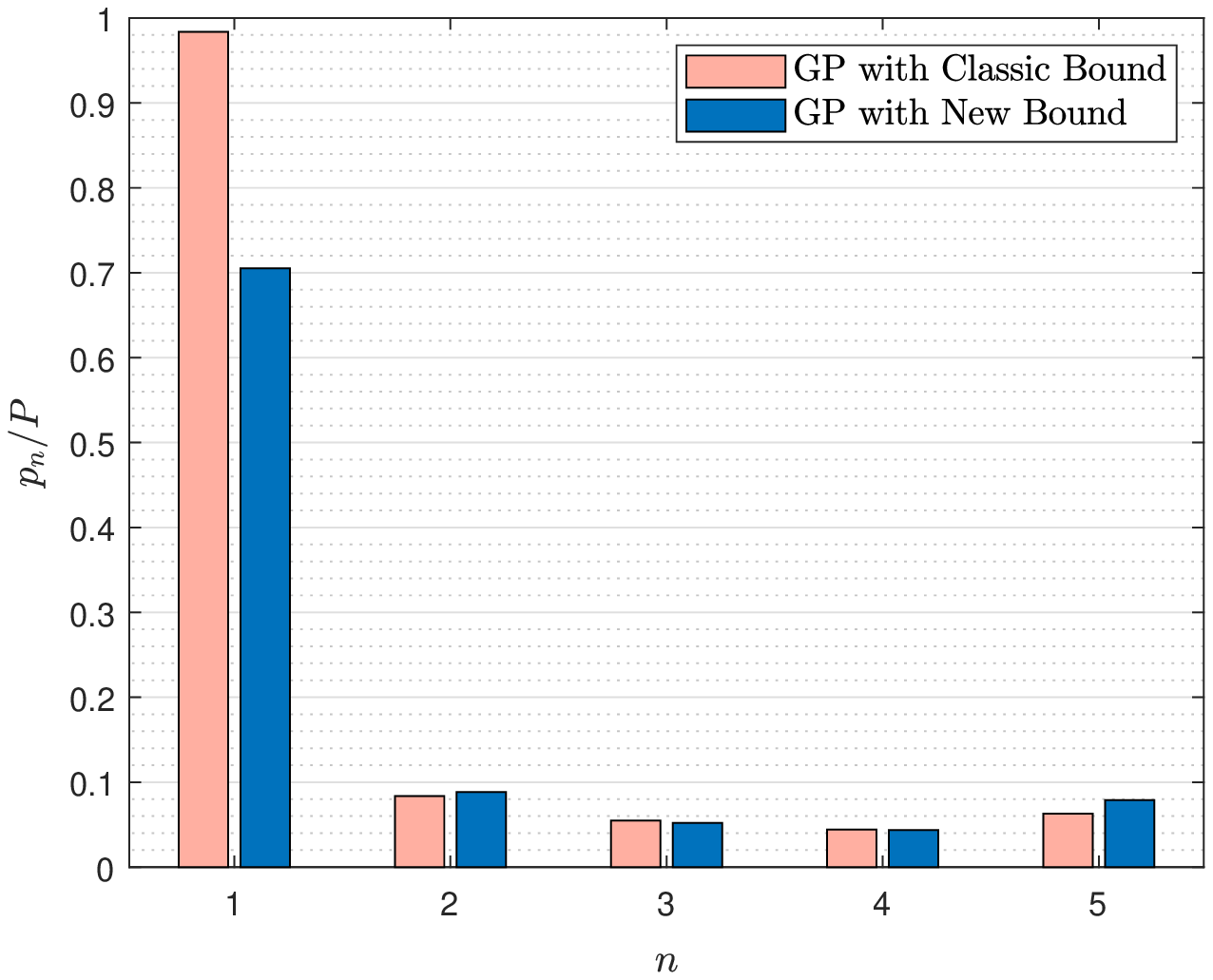}}
\vspace{-1em}
\caption{Transmit power in each round of HARQ for a SIMO channel.}
\label{fig:pn_4AT}
%\vspace{1em}
\end{minipage}
\begin{minipage}[b]{1.0\linewidth}
\centering
\centerline{\includegraphics[width=9.5cm]{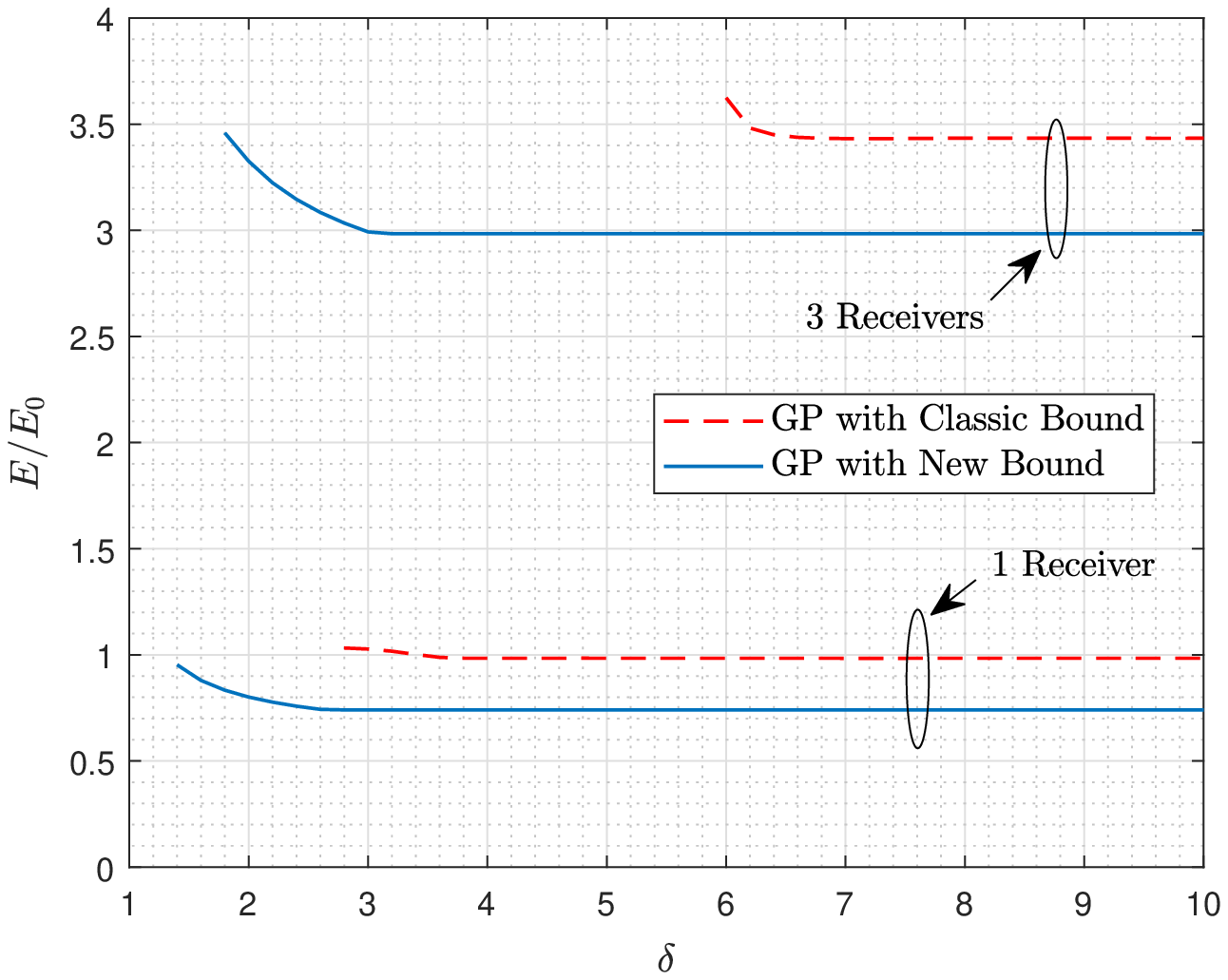}}
\vspace{-1em}
\caption{Energy vs. latency when $S=6$ and each receiver has 4 antennas.}
\label{fig:ED_4AT}
%\vspace{1em}
\end{minipage}
\end{figure}

We now take a closer look at the solutions obtained by GP assuming the two different outage probability bounds by comparing their respective power allocations across the blocks. Fig.~\ref{fig:pn_1AT} shows the transmit power in each round of HARQ for a single-input single-output (SISO) channel when $S=8$. As it turns out, GP with the new bound can save much energy mainly due to the fact that it sets a much lower (20\% less) power level for the initial transmission. Differing from the classic bound, the new bound puts more power in the subsequent retransmissions. This makes sense intuitively because the first block is always used whereas the retranmission blocks are used with rapidly decreasing probabilities. Notice that the two GP methods both use the max power level for the final block, which is used with the lowest probability among the five blocks. Moreover, if we remove the max power constraint and the latency constraint, and plot the resulting power decisions as in Fig.~\ref{fig:pn_1AT_unbounded}, then we see that the optimized solution would significantly raise $p_n$ especially at $n=5$; this result agrees with Remark \ref{remark:unbouned}.
As a further remark, note that $p_1>p_2$ in our optimized solution, whereas the true optimized power should increase in each retransmission as shown in \cite{Eriksson_14,Szczecinski_TCOM16}\footnote{But the exact value of the true optimized power is unknown in 
\cite{Eriksson_14,Szczecinski_TCOM16}.}. This is due to the approximation error $\hat Q_2>1$; similar errors can be observed in the earlier works \cite{Su_TCOM11} \cite{Ge_IET15} as well.

Next, we return to the latency constrained and power constrained setting, and plot the optimized power allocation for the $1\times4$ SIMO case in Fig.~\ref{fig:pn_4AT} for $S=8$. The figure shows that the transmit power levels all drop significantly by virtue of the multi-antenna diversity. But the power drops the least in the first block. This is because the initial transmission plays a key role in guaranteeing ultrareliability especially at $n=5$;y thus high transmit power must be secured for it. The new bound still sets lower power level than the classic bound in the first block.
Another observation worth noting from the above two figures is that the power curve is ``U''-shaped across the blocks, i.e., the initial and the final blocks tend to have higher power than the middle.

%============================

Moreover, Fig.~\ref{fig:ED_4AT} displays the tradeoff between energy and latency achieved by the two GP methods when $S=6$. Notice that the energy cost decreases when the latency constraint $\delta$ is relaxed. Observe also that the tradeoff curves all flatten out when $\delta$ is sufficiently large; in this regime the latency constraint does not have significant impact and the power decision only depends on the energy minimization objective and the target outage probability constraint. It can be seen that the classic bound is inferior to the new bound in two respects. First, the new bound enables GP to work under a much stricter latency constraint; second, at the same $\delta$, GP with the new bound yields lower energy consumption. The figure also shows that the performance gap between the two bounds is larger in the presence of multiple receivers.

%For ease of illustration, the power variables $p_n$ are normalized by the max power $P$.% so that $0\le p_n\le 1$. %The \documentclass[journal]{IEEEtran}

% Some Computer Society conferences also require the compsoc mode option,
% but others use the standard conference format.
%
% If IEEEtran.cls has not been installed into the LaTeX system files,
% manually specify the path to it like:
% \documentclass[conference]{../sty/IEEEtran}

%For ease of illustration, the power variables $p_n$ are normalized by the max power $P$.% so that $0\le p_n\le 1$. %The number of transmit antennas $L$, the average latency constraint $\delta$, and the pathloss-to-noise ratio $s$, are set differently in the following numerical examples.

\section{Conclusion}
\label{sec:conclusion}

The main contribution of this work is a novel upper bound on outage probability for HARQ with either incremental redundancy or Chase combining, which is much tighter than the classic bound when the transmit power are limited and hence is particularly suited for the IoT scenario. Using the new bound as a basis, we propose to approximate the intractable power control problem over multiple rounds of HARQ in a GP form that can be efficiently solved by convex optimization techniques. It is shown that the new bound and the resulting GP method are valid for a broad range of channel models. As shown in simulations, the proposed power control method can satisfy much stricter ultrareliability requirements while consuming much less energy than the state-of-the-art methods, because it approximates the problem more exactly by using the novel bound.

%By formulating the problem as a GP, we argue that the per-block power constraint, which is missing in some previous works, is necessary for ultrareliable communications. Further, the proposed upper bound along with the GP method for power control is extended to the multi-antenna channels in order to reap the antenna diversity gain.

\IEEEpeerreviewmaketitle

\bibliographystyle{IEEEbib}
\bibliography{IEEEabrv,strings}

\newpage
\begin{IEEEbiography}[{\includegraphics[width=1in,height=1.25in,clip,keepaspectratio]{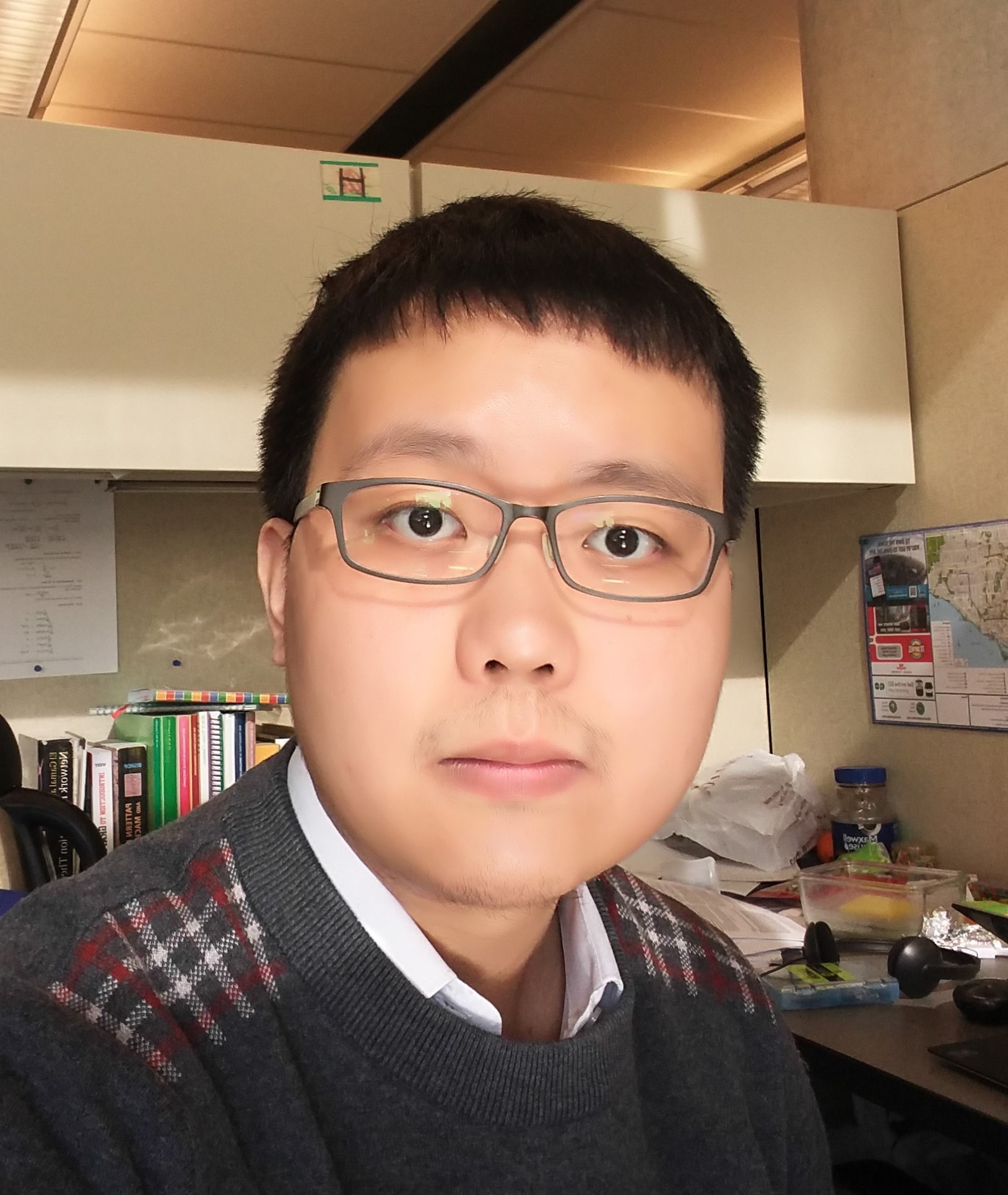}}]{Kaiming Shen}
(S'13-M'20) received the B.Eng. degree in information security and the B.S. degree in mathematics from Shanghai Jiao Tong University, Shanghai, China in 2011, then the M.A.Sc. and Ph.D. degrees in electrical and computer engineering from the University of Toronto, Ontario, Canada in 2013 and 2020, respectively.

Since 2020, he has been an Assistant Professor with the School of Science and Engineering at the Chinese University of Hong Kong (Shenzhen), China. His main research interests include optimization, wireless communications, data science, and information theory. Dr. Shen received the IEEE Signal
Processing Society Young Author Best Paper Award in 2021.
\end{IEEEbiography}

\begin{IEEEbiography}[{\includegraphics[width=1in,height=1.25in,clip,keepaspectratio]{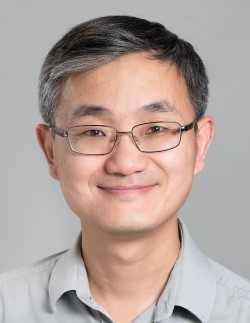}}]{Wei Yu}
(S'97-M'02-SM'08-F'14) received the B.A.Sc. degree in computer engineering and mathematics from the University of Waterloo, Waterloo, ON, Canada, in 1997, and the M.S. and Ph.D. degrees in electrical engineering from Stanford University, Stanford, CA, USA, in 1998 and 2002, respectively. 

Since 2002, he has been with the Electrical and Computer Engineering Department, University of Toronto, Toronto, ON, Canada, where he is currently a Professor and holds the Canada Research Chair (Tier 1) in Information Theory and Wireless Communications. He is a Fellow of the Canadian Academy of Engineering and a member of the College of New Scholars, Artists, and Scientists of the Royal Society of Canada. Prof. Wei Yu was the President of the IEEE Information Theory Society in 2021, and has served on its Board of Governors since 2015. He served as the Chair of the Signal Processing for Communications and Networking Technical Committee of the IEEE Signal Processing Society from 2017 to 2018. He was an IEEE Communications Society Distinguished Lecturer from 2015 to 2016. Prof. Wei Yu received the Steacie Memorial Fellowship in 2015, the IEEE Marconi Prize Paper Award in Wireless Communications in 2019, the IEEE Communications Society Award for Advances in Communication in 2019, the IEEE Signal Processing Society Best Paper Award in 2008, 2017 and 2021, the Journal of Communications and Networks Best Paper Award in 2017, and the IEEE Communications Society Best Tutorial Paper Award in 2015. He is currently an Area Editor of the IEEE Transactions on Wireless Communications, and in the past served as an Associate Editor for IEEE Transactions on Information Theory, IEEE Transactions on Communications, and IEEE Transactions on Wireless Communications.
\end{IEEEbiography}


\begin{IEEEbiography}[{\includegraphics[width=1in,height=1.25in,clip,keepaspectratio]{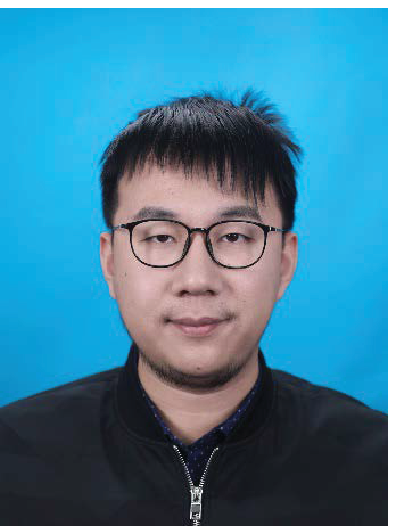}}]{Xihan Chen}
received the B.S. degree in electrical engineering from the Beijing University of Posts and Telecommunications, Beijing, China, in 2015, and the B.S. (Hons.) degree in electrical engineering from the Queen Mary University of London, London, U.K., in 2015. He was a visiting student in 2019 with the Department of Electronic and Computer Engineering, University of Toronto. He is currently working toward the Ph.D. degree at the College of Information Science \& Electronic Engineering, Zhejiang University, Hangzhou, China. His research interests include wireless communication and stochastic optimization.
\end{IEEEbiography}

\begin{IEEEbiography}[{\includegraphics[width=1in,height=1.25in,clip,keepaspectratio]{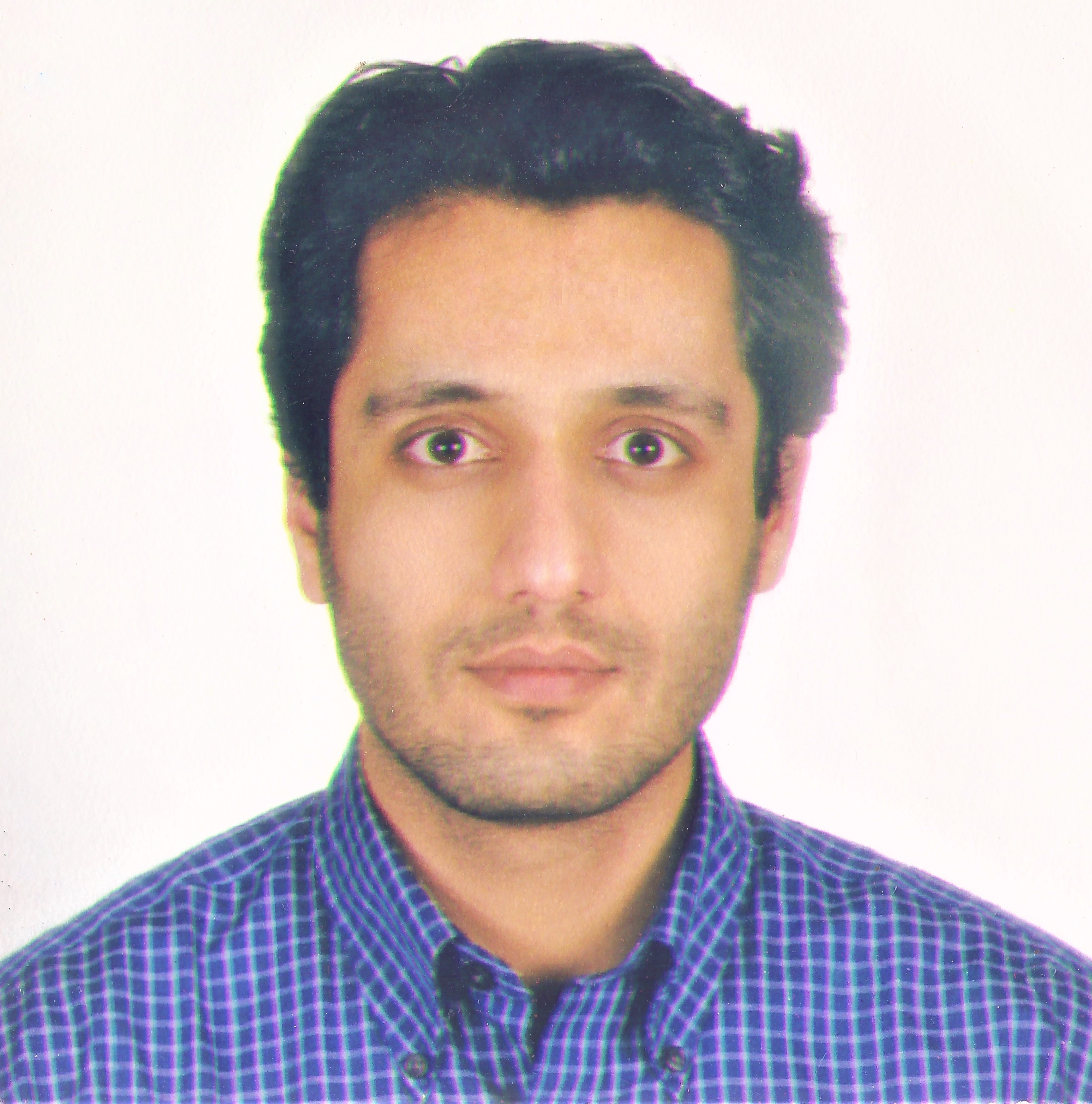}}]{Saeed R. Khosravirad}
 is a Member of Technical Staff at Nokia Bell Labs. In this role, he contributes to innovating the future generation of wireless networks with ultrareliable and low latency communications. He received his Ph.D. degree in telecommunications in 2015 from McGill University, Canada. Prior to that, he received the B.Sc. degree from the department of Electrical and Computer Engineering, University of Tehran, Iran, and the M.Sc. degree from the department of Electrical Engineering, Sharif University of Technology, Iran. During 2018-2019, he was with the Electrical \& Computer Engineering department of University of Toronto, Canada as a visiting scholar. He is an editor of the IEEE Transactions on Wireless Communications, and guest editor of the IEEE Wireless Communications magazine. His research fields of interest include wireless communications theory, ultrareliable communications for industrial automation, and wireless technologies for the beyond 5G era.
\end{IEEEbiography}

\end{document}